
\input harvmac
\input amssym.def
\input amssym
\baselineskip 14pt
\magnification\magstep1
\parskip 6pt
\newdimen\itemindent \itemindent=32pt
\def\textindent#1{\parindent=\itemindent\let\par=\resetpar%
\indent\llap{#1\enspace}\ignorespaces}

\let\oldpar=\par
\def\resetpar{\oldpar\parindent=20pt\let\par=\oldpar}

\font\ninerm=cmr9 \font\ninesy=cmsy9
\font\eightrm=cmr8 \font\sixrm=cmr6
\font\eighti=cmmi8 \font\sixi=cmmi6
\font\eightsy=cmsy8 \font\sixsy=cmsy6
\font\eightbf=cmbx8 \font\sixbf=cmbx6
\font\eightit=cmti8
\def\eightpoint{\def\rm{\fam0\eightrm}
  \textfont0=\eightrm \scriptfont0=\sixrm \scriptscriptfont0=\fiverm
  \textfont1=\eighti  \scriptfont1=\sixi  \scriptscriptfont1=\fivei
  \textfont2=\eightsy \scriptfont2=\sixsy \scriptscriptfont2=\fivesy
  \textfont3=\tenex   \scriptfont3=\tenex \scriptscriptfont3=\tenex
  \textfont\itfam=\eightit  \def\it{\fam\itfam\eightit}%
  \textfont\bffam=\eightbf  \scriptfont\bffam=\sixbf
  \scriptscriptfont\bffam=\fivebf  \def\bf{\fam\bffam\eightbf}%
  \normalbaselineskip=9pt
  \setbox\strutbox=\hbox{\vrule height7pt depth2pt width0pt}%
  \let\big=\eightbig  \normalbaselines\rm}
\catcode`@=11 %
\def\eightbig#1{{\hbox{$\textfont0=\ninerm\textfont2=\ninesy
  \left#1\vbox to6.5pt{}\right.\n@@space$}}}
\def\vfootnote#1{\insert\footins\bgroup\eightpoint
  \interlinepenalty=\interfootnotelinepenalty
  \splittopskip=\ht\strutbox %
  \splitmaxdepth=\dp\strutbox %
  \leftskip=0pt \rightskip=0pt \spaceskip=0pt \xspaceskip=0pt
  \textindent{#1}\footstrut\futurelet\next\fo@t}
\catcode`@=12 %
\def \d{{\rm d}}
\def \de{\delta}
\def \De{\Delta}

\def \nab{\nabla}
\def \pr{\partial}
\def \d{{\rm d}}

\def \hh{{\hat h}}
\def \hs{{\hat s}}
\def \hht{{\hat t}}

\def \bT{{\bar T}}

\def \bz{{\bar z}}
\def \bpsi{{\bar \psi}}
\def \bta{{\bar \eta}}

\def \l{\langle}
\def \r{\rangle}

\def \half{{\textstyle {1 \over 2}}}
\def \thir{{\textstyle {1 \over 3}}}
\def \quar{{\textstyle {1 \over 4}}}

\def \C{{\cal C}}

\def \E{{\cal E}}

\def \G{{\cal G}}

\def \N{{\cal N}}

\def \S{{\cal S}}

\def \tK{{\tilde K}}

\def \vphi{{\varphi}}
\def \oD{{\overline D}}

\font \bigbf=cmbx10 scaled \magstep1

\lref\hughtwo{J. Erdmenger and H. Osborn, Nucl. Phys. {B483} (1997)
431, hep-th/9605009.}
\lref\hughone{H. Osborn and A. Petkou,
    Ann. Phys. (N.Y.) {231} (1994) 311, hep-th/9307010.}
\lref\Symanzik{K. Symanzik, Lett. al Nuovo Cimento 3 (1972) 734.}
\lref\HO{H. Osborn, Ann. Phys. (N.Y.) 272 (1999) 243, hep-th/9808041.}
\lref\Erd{{\it Higher Transcendental Functions, vol. I}, ed. A. Erd\'elyi, 
(McGraw-Hill Book Co., New York, 1953), p. 222-242.}
\lref\Depp{M. D'Eramo, G. Parisi and L. Peliti, Lett. al Nuovo Cimento 2 (1971) 878.}
\lref\Lang{K. Lang and W. R\"uhl, Nucl. Phys. {B377} (1992) 371.}
\lref\LR{K. Lang and W. R\"uhl, Nucl. Phys. {B400} (1993) 597.}
\lref\LW{K. Lang and W. R\"uhl, Nucl. Phys. {B402} (1993) 573.}
\lref\Dot{Vl.S. Dotsenko, Nucl. Phys. B235 [FS11] (1984) 54.}
\lref\Uss{N.I. Ussyukina and A.I. Davydychev, Phys. Lett. B298 (1993) 363; 
B305 (1993) 136\semi
A.I. Davydychev and J.B. Tausk, Nucl. Phys. B397 (1993) 133.}
\lref\Pet{A.C. Petkou, Ann. Phys. (N.Y.) 249 (1996) 180, hep-th/9410093.}
\lref\Fone{S. Ferrara, A.F. Grillo, R. Gatto and G. Parisi, Nucl. Phys. 
B49 (1972) 77\semi
S. Ferrara, A.F. Grillo, R. Gatto and G. Parisi, Nuovo Cimento 19A
(1974) 667.}
\lref\Ftwo{S. Ferrara, A.F. Grillo and R. Gatto, Ann. Phys. 76 (1973) 161.}
\lref\Dob{V.K. Dobrev, V.B. Petkova, S.G. Petrova and I.T. Todorov,
Phys. Rev. D13 (1976) 887.}
\lref\West{E. D'Hoker, D.Z. Freedman, S.D. Mathur, A. Matusis,
L. Rastelli, {\it in} The Many Faces of the Superworld, ed. M.A. Shifman,
hep-th/9908160.\semi
B. Eden, P.S. Howe, C. Schubert, E. Sokatchev and P.C. West,
Phys. Lett. B472 (2000) 323, hep-th/9910150\semi
B. Eden, P.S. Howe, E. Sokatchev and P.C. West, hep-th/0004102\semi
M. Bianchi and S. Kovacs, Phys. Lett. B468 (1999) 102, hep-th/9910016\semi
J. Erdmenger and M. P\'erez-Victoria, Phys. Rev. D62 (2000) 045008,
Nucl. Phys. B589 (2000) 3, hep-th/9912250\semi
B. Eden, C. Schubert and E. Sokatchev, Phys. Lett. B482 (2000) 309, 
hep-th/0003096\semi
E. D'Hoker, J. Erdmenger, D.Z. Freedman and M. P\'erez-Victoria, 
hep-th/0003218.}
\lref\OPEC{G. Arutyunov, S. Frolov and A.C. Petkou, Nucl. Phys. B586 (2000) 
547, hep-th/0005182.}
\lref\Bel{A.A. Belavin, A.M. Polyakov and A.B. Zamolodchikov, Nucl. Phys.
B241 (1984) 333.}
\lref\one{F.A. Dolan and H. Osborn, Implications of $\N=1$ Superconformal
Symmetry for Chiral Fields, Nucl. Phys. B593 (2001) 599, hep-th/0006098.}
\lref\Grad{{\it Tables of Integrals, Series, and Products}, 5th ed., 
I.S. Gradshteyn and I.M. Ryzhik, ed. A. Jeffrey, 
(Academic Press, San Diego, 1994).}
\lref\Free{D.Z. Freedman, S.D. Mathur, A. Matsusis and L. Rastelli, 
Nucl. Phys. B546 (1999) 96, hep-th/9804058\semi
D.Z. Freedman, S.D. Mathur, A. Matsusis and L. Rastelli, 
Phys. Lett. B452 (1999) 61, hep-th/9808006\semi
E. D'Hoker and D.Z. Freedman, Nucl. Phys. B550 (1999) 261, hep-th/9811257\semi
E. D'Hoker and D.Z. Freedman, Nucl. Phys. B544 (1999) 612, hep-th/9809179.}
\lref\FreeI{E. D'Hoker, D.Z. Freedman and L. Rastelli, 
Nucl. Phys. B562 (1999) 395, hep-th/9905049.}
\lref\FreeD{E. D'Hoker, D.Z. Freedman, S.D. Mathur, A. Matsusis and 
L. Rastelli, Nucl. Phys. B562 (1999) 353, hep-th/9903196.}
\lref\Hok{E. D'Hoker, S.D. Mathur, A. Matsusis and L. Rastelli,
Nucl.Phys. B589 (2000) 38, hep-th/9911222.}
\lref\LiuT{H. Liu and A.A. Tseytlin, Phys. Rev. D59 (1999) 086002, 
hep-th/9807097.}
\lref\Liu{H. Liu, Phys. Rev. D60 (1999) 106005, hep-th/981152.}
\lref\San{S. Sanjay, Mod. Phys. Lett. A14 (1999) 1413, hep-th/990699.}
\lref\Brodie{J.H. Brodie and M. Gutperle, Phys Lett. B445 (1999) 296,
hep-th/9809067.}
\lref\Hoff{L.C. Hoffmann, A.C. Petkou and W. R\"uhl, 
Phys. Lett. B478 (2000) 320, hep-th/0002025\semi
L.C. Hoffmann, A.C. Petkou and W. R\"uhl, Adv. Theor. Math. Phys. 4 (2000),
hep-th/0002154.}
\lref\Frol{G. Arutyunov and S. Frolov, Phys. Rev. D62 (2000) 064016, 
hep-th/0002170.}
\lref\Herz{C.P. Herzog, JHEP 0102 (2001) 038, hep-th/0002039.}
\lref\Witt{E. Witten, Adv. Theor. Math. Phys. 2 (1998) 253, hep-th/9802150.}
\lref\HoffR{L. Hoffmann, L. Mesref and W. R\"uhl, Nucl. Phys. B589 (2000) 337.}
\lref\Eden{B. Eden, P.S. Howe, A. Pickering, E. Sokatchev and P.C. West,
Nucl. Phys. B581 (2000) 71, hep-th/0001138\semi
B. Eden, A.C. Petkou, C. Schubert and E. Sokatchev, Nucl. Phys., to
be published, hep-th/0009106.}
{\nopagenumbers
\rightline{DAMTP/00-125}
\rightline{hep-th/0011040}
\vskip 1.5truecm
\centerline {\bigbf Conformal Four Point Functions and the Operator Product
Expansion}
\vskip  6pt
\vskip 2.0 true cm
\centerline {F.A. Dolan and H. Osborn${}^\dagger$}

\vskip 12pt
\centerline {\ Department of Applied Mathematics and Theoretical Physics,}
\centerline {Silver Street, Cambridge, CB3 9EW, England}
\vskip 1.5 true cm

{\eightpoint
\parindent 1.5cm{

{\narrower\smallskip\parindent 0pt
Various aspects of the four point function for scalar fields in conformally
invariant theories are analysed. This depends on an arbitrary function of
two conformal invariants $u,v$.  A recurrence relation for the function 
corresponding to the contribution of an arbitrary spin field 
in the operator product expansion to the four point function
is derived. This is solved explicitly in two and four
dimensions in terms of ordinary hypergeometric functions of variables $z,x$
which are simply related to $u,v$. The operator product expansion analysis
is applied to the explicit expressions for the four point function found
for free scalar, fermion and vector field theories in four dimensions.
The results for four point functions obtained by using the AdS/CFT
correspondence are also analysed in terms of functions related to those
appearing in the operator product discussion.

PACS no: 11.25.Hf

Keywords: Conformal field theory, Operator product expansion, Four point
function

\narrower}}

\vfill
\line{${}^\dagger$ 
address for correspondence: Trinity College, Cambridge, CB2 1TQ, England\hfill}
\line{\hskip0.2cm emails:
{{\tt fad20@damtp.cam.ac.uk} and \tt ho@damtp.cam.ac.uk}\hfill}
}

\eject}
\pageno=1

\newsec{Introduction}

Much work has been undertaken in the last few years based on the AdS/CFT
correspondence in terms of understanding non trivial conformal field theories
in four, and also three and six, dimensions. In particular this has been
applied to $\N=4$ supersymmetric $SU(N)$ gauge theories when supergravity on
AdS${}_5$ determines the large $N$ limit of the associated conformal
field theory which is defined on the boundary. 

The correlation functions of operators on the boundary in the AdS/CFT
correspondence are then determined to leading order in $1/N$
by tree graphs with vertices given by the supergravity theory and with 
appropriate boundary/bulk and bulk/bulk propagators determined by the
Green functions on the AdS space. The form of the two and three point functions
are determined by conformal invariance while the four point function depends
on an arbitrary function of two conformal invariants. 

The four point function is of particular interest since it is constrained 
by the operator product expansion for any two fields. For
$\l \phi_1(x_1) \phi_2(x_2) \phi_3(x_3) \phi_4(x_4) \r$, which depends
on a function of two conformal invariant cross ratios $u,v$,  the operator
product expansion for $\phi_1(x_1) \phi_2(x_2)$ should provide in
principle an expansion for the four point function which is convergent
in some region when $x_1 \approx x_2$. Nevertheless this requires
knowing the form of the contribution, as a function of $u,v$, corresponding 
to operators of arbitrary spin, including all their derivatives, which 
is analogous to determining an explicit expression for
a partial wave expansion. In $d$ dimensions, when, with a Euclidean metric,
the conformal group is $O(d+1,1)$, then for scalar fields we are 
concerned just with the contribution of fields belonging to $(\ell,0,\dots)$
representations of $O(d)$, corresponding to fields 
$O^{(\ell)}{}_{\!\!\mu_1 \dots \mu_\ell}$
which are totally symmetric traceless rank $\ell$ tensors.  Although
conformal partial wave expansions were obtained long ago \refs{\Fone,\Dob,\LR}
they have not
been in a form which is easy to apply to disentangling the contributions
of different operators to the four point functions found through the 
AdS/CFT correspondence. This has necessitated approximate calculations of
just the leading terms in the power expansion for the contribution of 
operators with non zero spin in the operator product expansion \Herz.
More recently \OPEC\ results have been obtained for
the contributions of a conserved vector current and the energy momentum
tensor, which correspond to $\ell=1,2$ operators with dimensions $d-1,d$, to
the scalar four point function. We follow a similar approach to find
a recurrence relation for the contribution of operators for any $\ell$.
This recurrence relation may be solved when $d=2$, when the result follows
from the special simplifications arising from the use of complex 
coordinates, and also when $d=4$. In both cases the result is
expressible in terms of products of ordinary hypergeometric functions.
In two dimensions these depend on $\eta,\bta$ which are related to the
factors of $u,v$ found by using complex coordinates, $u=\eta\bta, \,
v=(1-\eta)(1-\bta)$ while in four dimensions we similarly define
$u=zx, \, v=(1-z)(1-x)$. The permutation symmetries of the four point
function may be translated into transformations of $z,x$ in addition
to the requirement of invariance under $z\leftrightarrow x$.

The structure of this paper is thus that in section 2 we review the
operator product expansion and obtain the recurrence relation for
the contribution of a quasi-primary operator of spin $\ell$ in $d$
dimensions to the scalar four point function in terms of the results for
$\ell-1, \, \ell-2$ by using the corresponding relation for Gegenbaur
polynomials. In section 3 we obtain compact explicit solutions for
arbitrary $\ell$ when $d=2,4$. In section 4 we briefly use these
results to obtain the complete contribution arising from the energy
momentum tensor, for which $\ell=2$. The overall normalisation is determined
by Ward identities. In section 5 we analyse the form of the integrals
arising in the AdS/CFT correspondence to the scalar four point function
in some simple cases. The results are related to a two variable function
$H$ introduced by us earlier \one\ which is related to the functions
arising in the operator product analysis. In section 6 we consider
the simple expressions for the four point function arising from free
conformal field theories and analyse their expansion in terms of the
conformal partial wave expressions obtained here for appropriate scale
dimension $\Delta$ and spin $\ell$. The results satisfy the positivity
conditions required by unitarity. Some mathematical details are deferred
to various appendices. In appendix A we construct the derivative operators
which appear in the operator product expansion for $\ell=1,2$ while in
appendix B we find, by direct calculation, the action of these differential
operators to give the contribution of an $\ell=1$ operator to the 
operator product expansion. The result is in accord with the less direct
discussion of section 2. In appendix C we describe some results for a function
$H$ which is obtained by the AdS/CFT integrals of section 5. This is
obtained explicitly in various cases of interest. In appendix D we sketch
some details of the derivation of the four point function for free vector
theories. 

\newsec{Operator Product Expansion Analysis}

For scalar operators $\phi_i$ of scale dimension $\De_i$ the contribution
of a spin $\ell$ operator $O^{(\ell)}{}_{\!\!\mu_1 \dots \mu_\ell}$ of
dimension $\Delta$ to
the operator product expansion, including all derivatives or descendants,
may be written as
\eqn\OPEpp{
\phi_1 (x_1) \phi_2(x_2) \sim
C_{\phi_1\phi_2O^{(\ell)}}\, {1\over r_{12}^{\, \, {1\over 2}(\De_1+\De_2 
-\Delta)}}\, C^{(\ell)}(x_{12}, \pr_{x_2})_{\mu_1 \dots \mu_\ell}
O^{(\ell)}{}_{\!\!\mu_1 \dots \mu_\ell} \, ,
}
where
\eqn\defxr{
x_{ij} = x_i - x_j \, , \qquad r_{ij} = (x_i - x_j)^2\, .
}
The derivative operator $C^{(\ell)}(s,\pr)$ is determined by the form
of the associated three and two point functions. For the former, conformal
invariance requires
\eqn\ppO{\eqalign{
\l & \phi_1 (x_1)  \phi_2(x_2) \, O^{(\ell)}(x_3) {\cdot \, \C} \r \cr
& {}= C_{\phi_1\phi_2O^{(\ell)}} \,
{1\over {r_{12}^{\,\,{1\over 2}(\De_1+\De_2 -\Delta + \ell)} \,
r_{13}^{\,\,{1\over 2}(\Delta + \De_{12} - \ell)} \,
r_{23}^{\,\,{1\over 2}(\Delta - \De_{12} - \ell)}}} \, 
Z_{\mu_1} \dots Z_{\mu_\ell} \C_{\mu_1 \dots \mu_\ell} \, , \cr}
}
where
\eqn\defZ{
Z_\mu = {x_{13\mu}\over r_{13}}-{x_{12\mu}\over r_{12}} \, , \qquad
Z^2 = {r_{12}\over r_{13}\, r_{23}} \, ,
}
transforms as a conformal vector at $x_3$ and $\C_{\mu_1 \dots \mu_\ell}$
is an arbitrary symmetric traceless tensor, $O^{(\ell)} {\cdot \C}
= O^{(\ell)}{}_{\!\!\mu_1 \dots \mu_\ell} \C_{\mu_1 \dots \mu_\ell}$,
and
\eqn\defdd{
\De_{ij} = \De_i - \De_j \, .
}
The two point function for $O^{(\ell)}$ is given by
\eqn\OO{
\l O^{(\ell)}(x_1) {\cdot \,\C} \, O^{(\ell)}(x_2) {\cdot \,\C}' \r =
{1\over r_{12}^{\, \, \Delta}} \, \C_{\mu_1 \dots \mu_\ell}
I_{\mu_1 \nu_1}(x_{12}) \dots I_{\mu_\ell \nu_\ell}(x_{12})\,
\C'{}_{\! \nu_1 \dots \nu_\ell} \, ,
}
where
\eqn\defI{
I_{\mu\nu}(x) = \de_{\mu\nu} - 2 {x_\mu x_\nu \over x^2} \, ,
}
is the inversion tensor. For completeness we assume the normalisation of
the scalar fields is determined by
\eqn\pp{
\l \phi_i(x_1) \phi_j (x_2) \r = \de_{ij} {1\over r_{12}^{\, \, \De_i}} \, .
}

As a consequence of \ppO\ and \OO\ we must therefore require in \OPEpp\
$C^{(\ell)}(s,\pr) = C^{{1\over 2}(\Delta + \De_{12} - \ell),
{1\over 2}(\Delta - \De_{12} - \ell)}(s,\pr)$ where
\eqn\defC{\eqalign{
C^{a,b}(x_{12}, \pr_{x_2})_{\mu_1 \dots \mu_\ell}&
{1\over r_{23}^{\, \, S}} I_{\mu_1 \nu_1}(x_{23}) \dots 
I_{\mu_\ell \nu_\ell}(x_{23})\, \C_{\! \nu_1 \dots \nu_\ell} \cr
&{}= {1\over r_{13}^{\, \, a} \, r_{23}^{\, \, b}}\,
Z_{\mu_1} \dots Z_{\mu_\ell} \C_{\mu_1 \dots \mu_\ell} \, ,\cr}
}
and
\eqn\Sab{
S=a+b+\ell \, .
}
We construct $C^{a,b}(s,\pr)$ explicitly in appendix A for $\ell=1,2$
where it is given in terms of the known results for $\ell=0$. The
generalisation to arbitrary $\ell$ is evident but is not needed here.

If \OPEpp\ is applied in the four point function then the corresponding
contribution has the form
\eqn\OPEpppp{\eqalign{
\l \phi_1(x_1)& \phi_2(x_2) \phi_3(x_3) \phi_4(x_4) \r \cr
\sim {}& 
{1\over r_{12}^{\,\,{1\over 2}(\De_1+\De_2)}
r_{34}^{\,\,{1\over 2}(\De_3+\De_4)}}
\bigg ({r_{24}\over r_{14}} \bigg )^{\! {1\over 2}\De_{12}} \!
\bigg ({r_{14}\over r_{13}} \bigg )^{\! {1\over 2}\De_{34}} \!\!\!
C_{\phi_1\phi_2O^{(\ell)}} C_{\phi_3\phi_4O^{(\ell)}}\cr
&{}\times u^{{1\over 2}(\Delta-\ell)} 
G^{(\ell)}\big (\half(\Delta - \De_{12} - \ell),
\half(\Delta + \De_{34} - \ell),\Delta;u,v\big ) \, , \cr}
}
which depends on the two conformal invariants
\eqn\defuv{
u= {r_{12} \, r_{34} \over r_{13} \, r_{24}} \, , \qquad \qquad
v= {r_{14} \, r_{23} \over r_{13} \, r_{24}} \, .
}
The functions $G^{(\ell)}$ are determined by
\eqn\defGl{
C^{a,b}(x_{12}, \pr_{x_2})_{\mu_1 \dots \mu_\ell}{1\over 
r_{23}^{\,\,e}\, r_{24}^{\,\,f}}Y_{\mu_1}\dots Y_{\mu_\ell}
= {1\over r_{14}^{\,\,a}\, r_{24}^{\,\,b}}
\bigg ({r_{14}\over r_{13}} \bigg )^{\! e}
G^{(\ell)} (b,e,S;u,v) \, ,
}
where
\eqn\defY{
Y_\mu = {x_{32\mu}\over r_{23}}-{x_{42\mu}\over r_{24}} \, , 
}
and, in addition to \Sab, we have
\eqn\Sef{
S=e+f+\ell \, .
}

We have undertaken a direct evaluation of \defGl\ in appendix B for $\ell=1$
but for a discussion of arbitrary $\ell$ we
establish a recurrence relation which may be used iteratively to determine
$G^{(\ell)}$ from $G^{(0)}$. We start from the integral representation
\eqn\intrep{\eqalign{
\C_{\mu_1 \dots \mu_\ell}& 
{1\over r_{23}^{\,\,e}\, r_{24}^{\,\,f}}Y_{\mu_1}\dots Y_{\mu_\ell}\cr
={}& r_{34}^{\,\,{1\over 2}d-S} 
\C_{\mu_1 \dots \mu_\ell} \, N_{\ell,e,f} \int \!\! \d^d x \,
{1\over (x_2-x)^{2S} (x_3-x)^{2({1\over 2}d - f - \ell)}
(x_4-x)^{2({1\over 2}d - e - \ell)}}\cr
& \qquad \qquad \qquad \qquad\quad\times 
I_{\mu_1 \nu_1}(x_2-x) \dots I_{\mu_\ell \nu_\ell}(x_2-x)
X'{}_{\! \nu_1} \dots X'{}_{\! \nu_\ell} \, ,\cr}
}
with
\eqn\defXp{
X' = {x_3-x \over (x_3-x)^2} - {x_4-x \over (x_4-x)^2} \, .
}
The general structure of \intrep\ follows from conformal invariance
assuming \Sef\ and is a generalisation of the well known result when
$\ell=0$ \Depp. To obtain the overall constant in \intrep\ we may define
\eqn\defy{
y= {x-x_2 \over (x-x_2)^2} \, , \qquad
y_i = {x_i-x_2 \over (x_i-x_2)^2} \, , \ i=2,3 \, , \qquad
Y' = {y_3 - y \over (y_3-y)^2} - {y_4 - y \over (y_4-y)^2} \, ,
}
so that, for $\alpha_2+\alpha_3+\alpha_4+\ell = d$,
\eqnn\intxy$$\eqalignno{
\C_{\mu_1 \dots \mu_\ell} \int & \! \d^d x \,
{1\over (x_2-x)^{2\alpha_2} (x_3-x)^{2\alpha_3} (x_4-x)^{2\alpha_4}}
I_{\mu_1 \nu_1}(x_2-x) \dots I_{\mu_\ell \nu_\ell}(x_2-x)
X'{}_{\! \nu_1} \dots X'{}_{\! \nu_\ell} \cr
= {}& y_3^{\,\, 2\alpha_3} y_4^{\,\, 2\alpha_4}
\C_{\mu_1 \dots \mu_\ell} \int \! \d^d y \, 
{1\over (y_3-y)^{2\alpha_3} (y_4-y)^{2\alpha_4}}
Y'{}_{\! \mu_1} \dots Y'{}_{\! \mu_\ell} \cr
= {}& \pi^{{1\over 2}d} \prod_{r=0}^{\ell-1} (d-1-\alpha_2+r)
\prod_{i=1}^3 {\Gamma(\half d - \alpha_i) \over \Gamma(\alpha_i+\ell)}\cr
&{}\times 
\C_{\mu_1 \dots \mu_\ell} (y_3 - y_4)_{\mu_1} \dots (y_3 - y_4)_{\mu_\ell}
{y_3^{\,\, 2\alpha_3} y_4^{\,\, 2\alpha_4} \over (y_3-y_4)^{2({1\over 2}d -
\alpha_2)}} \, . & \intxy \cr}
$$
It is easy to see that \intxy\ is in accord with \intrep\ if we take
\eqn\Nr{
N_{\ell,e,f} = {1\over \pi^{{1\over 2}d}}\,
{1\over (d -1 - S)_\ell} \, {\Gamma(\half d - e)
\Gamma(\half d - f) \Gamma(S+\ell) \over \Gamma(e+\ell) \Gamma(f+\ell)
\Gamma(\half d - S )} \, , 
}
for
\eqn\Poch{
(\gamma)_n = {\Gamma(\gamma+n) \over \Gamma(\gamma)} \, .
}

We may now use \intrep\  in \defGl\ and apply the definition \defC\ 
to give
\eqn\Cx{\eqalign{
C^{a,b}(x_{12}, &\pr_{x_2})_{\mu_1 \dots \mu_\ell}{1\over (x_2-x)^{2S}}
I_{\mu_1 \nu_1}(x_2-x) \dots I_{\mu_\ell \nu_\ell}(x_2-x)
X'{}_{\! \nu_1} \dots X'{}_{\! \nu_\ell}\cr
{}& = {1\over (x_1-x)^{2a} (x_2-x)^{2b}}X_{\mu_1} \dots X_{\mu_\ell}
\E^{(\ell)}{}_{\!\! \mu_1 \dots \mu_\ell, \nu_1 \dots \nu_\ell}
X'{}_{\! \nu_1} \dots X'{}_{\! \nu_\ell} \, ,\cr}
}
where $\E^{(\ell)}{}_{\!\! \mu_1 \dots \mu_\ell, \nu_1 \dots \nu_\ell}$ is
the projector onto symmetric traceless rank $\ell$ tensors, and
\eqn\defX{
X = {x_1-x \over (x_1-x)^2} - {x_2-x \over (x_2-x)^2} \, .
}
The contraction in \Cx\ may further be evaluated as
\eqn\Geg{\eqalign{
X_{\mu_1} \dots X_{\mu_\ell}
\E^{(\ell)}{}_{\!\! \mu_1 \dots \mu_\ell, \nu_1 \dots \nu_\ell}
X'{}_{\! \nu_1} \dots X'{}_{\! \nu_\ell} ={}& {\ell ! \over 2^\ell
(\half d-1)_\ell} \big ( X^2 X'{}^2 \big )^{{1\over 2}\ell}
C^{{1\over 2}d-1}_\ell ( t) \, , \cr 
t= {}& {X{\cdot X'} \over \big ( X^2 X'{}^2 \big )^{1\over 2}} \, .\cr}
}
$C^\lambda_n(t)$ are Gegenbaur polynomials of order $\ell$
(for $\lambda=\half$ these are just Legendre polynomials). In consequence
\defGl\ gives
\eqnn\Gint
$$\eqalignno{
r_{34}^{\,\,{1\over 2}d-S} &
{\ell ! \over 2^\ell (\half d-1)_\ell}N_{\ell,e,f} 
\int \! \d^d x \, {\big ( X^2 X'{}^2 \big )^{{1\over 2}\ell}
C^{{1\over 2}d-1}_\ell (t)
\over (x_1-x)^{2a} (x_2-x)^{2b} (x_3-x)^{2({1\over 2}d - f - \ell)}
(x_4-x)^{2({1\over 2}d - e - \ell)}} \cr
& {} =  {1\over r_{14}^{\,\,a}\, r_{24}^{\,\,b}}
\bigg ({r_{14}\over r_{13}} \bigg )^{\! e} \Big (
G^{(\ell)} (b,e,S;u,v) \cr
&\qquad \qquad \qquad \qquad {} + K^{(\ell)}_{b,e,S} u^{{1\over 2}d-S}
G^{(\ell)}(\half d + b-S, \half d + e - S,d-S;u,v) \Big ) \, . & \Gint \cr}
$$
The last term on the right hand side of \Gint\ is a so called shadow
term which is non analytic at $u=0$.\foot{Both terms in \Gint\ satisfy
the recurrence relation derived for $G^{(\ell)}$ below. Using this
result and the evaluation of \Gint\ for $\ell=0$ determines
$$\eqalign{
K^{(\ell)}_{b,e,S} = {}&{\Gamma(\half d+b-S+\ell)\Gamma(\half d+e-S+\ell)
\Gamma(S+\ell)\Gamma(S+\ell-1)
\over \Gamma(b+\ell) \Gamma(e+\ell)}\cr
&{}\times
{\Gamma(\half d -b) \Gamma(\half d -e) \Gamma(S-\half d)\Gamma(d-S-1)\over
\Gamma(S-b) \Gamma(S-e) \Gamma(\half d -S) \Gamma(S-1)}\, .\cr}
$$}
This term may be neglected in our 
analysis of $G^{(\ell)}$.
To evaluate the left hand side of \Gint\  we may note that
\eqn\XXp{
X^2 = {r_{12}\over (x_1-x)^2(x_2-x)^2} \, , \qquad \qquad
X'^2 = {r_{34}\over (x_3-x)^2(x_4-x)^2} \, ,
}
and
\eqn\Xpr{\eqalign{
2X{\cdot X'} = {}& - {r_{13}\over (x_1-x)^2(x_3-x)^2} + 
{r_{23}\over (x_2-x)^2(x_3-x)^2}\cr
{}& - {r_{24}\over (x_2-x)^2(x_4-x)^2} +
{r_{14}\over (x_1-x)^2(x_4-x)^2} \, ,\cr}
}
so that the integral in \Gint\ is reducible to linear combinations of the 
form
\eqn\fourp{
\int \! \d^d x \, {1
\over (x_1-x)^{2\alpha_1} (x_2-x)^{2\alpha_2} (x_3-x)^{2\alpha_3}
(x_4-x)^{2\alpha_4}} \, , \quad \sum_i \alpha_i = d \, ,
}
which are expressible in terms of functions of the conformal invariants
$u,v$ defined in \defuv.
A useful recurrence relation may be obtained by using the recurrence
relation for Gegenbaur polynomials,
\eqn\recur{
\ell C^\lambda_\ell(t) = 2(\lambda+\ell-1) t C^\lambda_{\ell-1}(t)
- (2\lambda + \ell -2) C^\lambda_{\ell-2}(t) \, .
}
Substituting this into \Gint\ and with \Nr\ we may then obtain using
\XXp\ and \Xpr\ a corresponding relation for $G^{(\ell)}$,
\eqnn\recurG
$$\eqalignno{
& G^{(\ell)} (b,e,S;u,v) \cr
{}& = {1\over 2} {S+\ell-1 \over d-S+\ell-2}
\bigg \{ {\half d - e -1 \over f + \ell -1} \Big (
v G^{(\ell-1)} (b+1,e+1,S;u,v) - G^{(\ell-1)} (b,e+1,S;u,v) \Big ) \cr
& \qquad \qquad \qquad \qquad {}+ {\half d - f -1 \over e + \ell -1} \Big (
G^{(\ell-1)} (b,e,S;u,v) - G^{(\ell-1)} (b+1,e,S;u,v) \Big ) \bigg \} \cr
& {} - {1\over 4} {(S+\ell-1)(S+\ell-2) \over (d-S+\ell-2)(d-S+\ell-3)}\,
{(\half d - e -1)(\half d - f -1) \over (f + \ell -1)(e + \ell -1)}\,
{(\ell-1)(d+\ell-4)\over (\half d +\ell -2)(\half d +\ell - 3)} \cr
& \quad {} \times u G^{(\ell-2)} (b+1,e+1,S;u,v) \, , & \recurG \cr}
$$
where $f$ is determined by \Sef.

The starting point in the iteration $G^{(0)}$, corresponding to a scalar
field in the operator product expansion, has been obtained by various
authors. In the short distance limit $x_1 \to x_2$ we have $u\to 0, \, v\to 1$
and it is given as a power series in $u, \, 1-v$,
\eqn\Gzero{
G^{(0)} (b,e,S;u,v) = G(b,e,S+1-\half d, S;u,1-v) \, ,
}
where
\eqn\defG{\eqalign{
G(\alpha,\beta,\gamma,\delta;u,1&{} -v) =  
G(\beta,\alpha,\gamma,\delta;u,1 -v) \cr
={}& \sum_{m,n=0} {(\delta-\alpha)_m
(\delta-\beta)_m \over m! (\gamma)_m} \,
{(\alpha)_{m+n} (\beta)_{m+n} \over n! (\delta)_{2m+n}} \, u^m (1-v)^n \cr
= {}& {\Gamma(\delta) \Gamma(\delta-\alpha-\beta)\over \Gamma(\delta-\alpha)
\Gamma(\delta-\beta)} \, F_4(\alpha,\beta,\gamma,\alpha+\beta+1-\delta;u,v) \cr
{} +{}& {\Gamma(\delta) \Gamma(\alpha+\beta-\delta)\over \Gamma(\alpha)
\Gamma(\beta)} \, v^{\delta-\alpha-\beta}
F_4(\delta-\alpha,\delta-\beta,\gamma,\delta-\alpha-\beta+1;u,v) \,,\cr}
}
with $F_4$ one of the well documented Appell functions, ref.\Grad, 
p. 1080-1084.  Two important symmetry
relations may be obtained from the associated result demonstrated for $G$ 
in \one\ and checking
consistency with \recurG. These correspond to letting $x_1 \leftrightarrow
x_2, \, a \leftrightarrow b$ and also $x_3 \leftrightarrow x_4, \,
e \leftrightarrow f$, and using \Sab\ and \Sef\ we have
\eqn\symG{\eqalign{
G^{(\ell)} (b,e,S;u,v) = {}& (-1)^\ell v^{-e} G^{(\ell)} (a,e,S;u',v')\, ,\cr
 = {}& (-1)^\ell v^{-b} G^{(\ell)} (b,f,S;u',v') \, , \qquad u'=u/v \, , \
v' = 1/v \, . \cr}
}

Although \recurG\ is rather involved it may be solved directly when
$u=0$ when the last term is absent. From \Gzero\ and \defG\ it is evident
that $G^{(0)} (b,e,S;0,v) = F(b,e;S;1-v)$, an ordinary hypergeometric
function and by using standard hypergeometric identities
\eqn\uzero{
G^{(\ell)} (b,e,S;0,v) = \big ( {- \half (1-v)} \big)^\ell
F(b+\ell,e+\ell;S+\ell;1-v ) \, .
}
Other general results for arbitrary $d$ are hard to find but for $v=1$
we may determine the leading behaviour for small $u$,
\eqn\vone{
G^{(\ell)} (b,e,S;u,1) \sim \cases{{\displaystyle 
{1\over 2^\ell}(-1)^{{1\over 2}\ell}
g_\ell \, u^{{1\over 2}\ell}} \, , & $\ell$ even, \cr
{\displaystyle
{1\over 2^{\ell+1}}(-1)^{{1\over 2}(\ell-1)}g_{\ell+1}
{(\ell + \half d -1) (a-b)(e-f)\over(S-d-\ell+2)(S+\ell)}\,
u^{{1\over 2}(\ell+1)}} \, , & $\ell$ odd, \cr}
}
where from \recurG\ $g_\ell=(\ell-1)(\ell+d-4)g_{\ell-2}/(\ell+\half d-2)
(\ell+\half d-3)$ and starting from $g_0=1$,
\eqn\gl{
g_\ell={\ell ! \, (\half d-1)_{{1\over 2}\ell} \over (\half \ell)! \,
(\half d-1)_\ell} \, .
}  

\newsec{Solutions in Two and Four Dimensions}

We show here how the recursion relation \recurG\ may be solved explicitly
in two and four dimensions. 

For $d=2$, $G^{(\ell)}$ may be found directly
using the simplifications obtained through using complex coordinates $z,\bz$
in this case. Letting $x^z \equiv z$ and, from $x^2 = z \bz$, $x_z = \half \bz$
then the inversion tensor defined in \defI\ is, in this complex basis,
$I_{z\bz}(x) = I_{\bz z}(x)=0 $ and $I_{zz}(x)= - \half \bz/z, \
I_{\bz\bz}(x)= - \half z/\bz$. Since \defZ\ reduces to
\eqn\Ztwo{
Z^z = - {\bz_{12} \over \bz_{13}\, \bz_{23}} \, , \qquad
Z^\bz = - {z_{12} \over z_{13}\, z_{23}} \, \,
}
the equation \Cx\ becomes
\eqn\Ctwo{
C^{a,b}(x_{12},\pr_{x_2})^{z\dots z} 
{1\over z_{23}^{\,\, S+\ell} \bz_{23}^{\,\, S-\ell}}
= {z_{12}^{\,\, \ell} \over z_{13}^{\,\, a+\ell} z_{23}^{\,\, b+\ell}}\,
{1\over \bz_{13}^{\,\, a} \, \bz_{23}^{\,\, b}} \, ,
}
together with its conjugate. The differential operator therefore
factorises in the form\Bel
\eqn\Cfact{
C^{a,b}(x_{12},\pr_{x_2})^{z\dots z} = z_{12}^{\,\, \ell} \,
{}_1F_1(a+\ell,S+\ell;z_{12}\pr_{z_2})
{}_1F_1(a,S-\ell;\bz_{12}\pr_{\bz_2}) \, ,
}
where ${}_1F_1(\alpha,\beta;x) = \sum_n (\alpha)_n /n!(\beta)_n \, x^n$.
Since
\eqn\Ft{
{}_1\! F_1(a;S;z_{12} \pr_{z_2}) {1\over z_{23}^{\ \,e} \, z_{24}^{\ \,f}}
= {1\over z_{14}^{\ \,a}\, z_{24}^{\ \,b}}
\Big ( {z_{14}\over z_{13}} \Big )^{\! e} F(b,e;S;\eta) \, , \quad
S=a+b=e+f \, ,
}
where
\eqn\defeta{
\eta = {z_{12}\, z_{34} \over z_{13} \, z_{24}} \, ,
}
the definition \defGl\ gives
\eqn\Gtwo{\eqalign{
G^{(0)}(b,e,S;u,v)= {}& F(b,e;S;\eta) F(b,e;S;\bta) \, , \cr
G^{(\ell)}(b,e,S;u,v)={}& (-\half \eta)^\ell 
F(b+\ell,e+\ell;S+\ell;\eta) F(b,e;S-\ell;\bta) \cr
&{} + \hbox{conjugate} \, , \quad  \ell >0 \, , \cr}
}
for
\eqn\uvc{
u = \eta \bta \, , \qquad v = (1-\eta)(1-\bta) \, .
}

For $G^{(0)}$ the validity of the result \Gtwo\ depends, from \Gzero, on 
the factorisation formula shown in \one\ for the function $G$ defined
in \defG. This is obtained from a similar reduction formula for $F_4$
and takes the form
\eqn\redG{
G(\alpha,\beta,\gamma,\gamma;u,1-v)= F(\alpha,\beta;\gamma;x)
F(\alpha,\beta;\gamma;z) \, .
}
where
\eqn\uvxz{
u = xz \, , \qquad v=(1-x)(1-z) \, .
}
We have verified that $G^{(\ell)}$ defined then through \recurG\ for $\ell
= 1,2,\dots $ agrees with \Gtwo. The calculation is very similar to the
$d=4$ case which we describe next.

When $d=4$ the result for $G^{(0)}$ given by \Gzero\ and \defG\ may
also be simplified by use of a reduction formula extending \redG\ 
which was also obtained in \one,
\eqn\redGG{\eqalign{\!\!\!\!\!
G(\alpha,\beta,\gamma ,\gamma+1;u,1-v)
=  {1\over z-x}& \big (
z F(\alpha-1,\beta-1;\gamma-1;x)F(\alpha,\beta;\gamma+1;z) \cr
{}&
- x F(\alpha,\beta;\gamma+1;x)F(\alpha-1,\beta-1;\gamma-1;z) \big )\, . \cr}
}
The solution for any $\ell$ which follows from this is then
\eqn\Glfour{\eqalign{
G^{(\ell)}(b,e,S;u,v)= {}& {1\over z-x} \Big ((-\half z)^\ell
z F(b-1,e-1;S-2-\ell;x)F(b+\ell,e+\ell;S+\ell;z) \cr
&\qquad {} - (-\half x)^\ell
x F(b-1,e-1;S-2-\ell;z)F(b+\ell,e+\ell;S+\ell;x) \Big )\, . \cr}
}
The verification of \Glfour\ from \recurG\ depends on
\eqnn\rrr
$$\eqalignno{
z & F(b-1,e-1;S-2-\ell;x)F(b+\ell,e+\ell;S+\ell;z)\cr
{}& = - {S+\ell-1 \over S-\ell-2}
\bigg \{ {e -1 \over f + \ell -1} \Big (
(1-z)(1-x) F(b,e;S-1-\ell;x)F(b+\ell,e+\ell;S+\ell-1;z) \cr
& \qquad \qquad \qquad \qquad \qquad {} -
F(b-1,e;S-1-\ell;x)F(b+\ell-1,e+\ell;S+\ell-1;z) \Big )\cr
& \qquad \qquad {}+ {f -1 \over e + \ell -1} \Big (
F(b-1,e-1;S-1-\ell;x)F(b+\ell-1,e+\ell-1;S+\ell-1;z) \cr
& \qquad \qquad \qquad \qquad \qquad {} -
F(b,e-1;S-1-\ell;x)F(b+\ell,e+\ell-1;S+\ell-1;z) \Big ) \bigg \} \cr
& \quad {} - {(S+\ell-1)(S+\ell-2) \over (S-\ell-2)(S-\ell-1)}\,
{(e -1)(f -1) \over (f + \ell -1)(e + \ell -1)} \cr
& \qquad \qquad {} \times  
x F(b,e;S-\ell;x)F(b+\ell-1,e+\ell-1;S+\ell-2;z) \, .  & \rrr \cr}
$$
This corresponds exactly to the form of \recurG\ for $d=4$ except
if $\ell=1$ when the last term, which matches the last term in \rrr,
is missing. However in this case this piece  times $z$
is symmetric under $z\leftrightarrow x$ and it is then cancelled
by the other term in \Glfour\ which is obtained from the analogous result to \rrr\
for $z\leftrightarrow x$. The justification of \rrr\ depends on
standard hypergeometric identities, ref.\Grad, p. 1071, in particular we use,
\eqn\hrel{\eqalign{
\!\!\!\!(S+\ell-1)&\big ( (1-z) F(b+\ell,e+\ell;S+\ell-1;z) -
F(b+\ell-1,e+\ell;S+\ell-1;z) \big ) \cr
{}& = - (f + \ell -1) z F(b+\ell,e+\ell;S+\ell;z) \, , \cr
\!\!\!\!(S+\ell-1)&\big ( F(b+\ell,e+\ell-1;S+\ell-1;z) -
F(b+\ell-1,e+\ell-1;S+\ell-1;z) \big ) \cr
{}& =  (e + \ell -1) z F(b+\ell,e+\ell;S+\ell;z) \, , \cr
\!\!\!\!(e-1)(1&{}-x)F(b,e;S-1-\ell;x) + (f-1) F(b+\ell,e+\ell-1;S+\ell-1;z)\cr
{}& = (S-\ell-2) F(b-1,e-1;S-2-\ell;x) \, . \cr}
}

For consistency we may check that \Glfour\ satisfies the consistency 
relations \symG. It is crucial to recognise that \uvxz\ is invariant
under $x\leftrightarrow z$, under which of course \Glfour\ is invariant.
For the transformations corresponding to  $x_1 \leftrightarrow
x_2$ or $x_3 \leftrightarrow x_4$ we choose
\eqn\txz{
x \to x' = {x\over x-1} \, , \quad
z \to z' = {z\over z-1}  \quad \Rightarrow \quad u \to u' = {u\over v} \, ,
\quad v\to v' ={1\over v} \, .
}
Using standard hypergeometric results, ref.\Grad, p. 1069, 
with \Sab, \Sef\ we have
\eqn\tGf{\eqalign{
F(b-1,e-1;S-2-\ell;x) = {}& (1-x)^{-e+1}F(a-1,e-1;S-2-\ell;x')\cr 
= {}& (1-x)^{-b+1} F(b-1,f-1;S-2-\ell;x')\, , \cr
F(b+\ell,e+\ell;S+\ell;z) = {}& (1-z)^{-e-\ell} F(a+\ell,e+\ell;S+\ell;z')\cr
= {}& (1-z)^{-b-\ell} F(b+\ell,f+\ell;S+\ell;z') \, . \cr}
}
which in \Glfour, with $z-x=-(z'-x')v$, are sufficient to verify \symG.

Both results \Gtwo\ and \Glfour\ are of course compatible with
$G^{(0)}(0,0,0;u,v)=1$, representing the contribution of the identity
operator in the operator product expansion.

\newsec{Energy Momentum Tensor}

It is of interest to specialise the general discussion to the 
particular case of the energy momentum tensor $T_{\mu\nu}$ which is
an $\ell=2$ operator with dimension $d$ satisfying the conservation
equation $\pr_\mu T_{\mu\nu}=0$. In this case the normalisation is not
fixed by \OO\ but instead through Ward identities. For the canonically
normalised energy momentum tensor, but with the scalar field $\phi$
still normalised as in \pp, the $\l \phi \phi T\r$ three point function
obtained from \ppO\ becomes \hughone
\eqn\ppT{
\l \phi (x_1) \phi(x_2) \, T_{\mu\nu}(x_3) \r 
= - {1\over S_d} \, {\De d\over d-1} \,
{1\over {r_{12}^{\,\,\,\De -{1\over 2}d}} \,
r_{13}^{\,\,{1\over 2}d} \, r_{23}^{\,\,{1\over 2}d}} \,
\bigg ( {Z_{\mu} Z_{\nu}\over Z^2} - {1\over d} \, \de_{\mu\nu} \bigg ) \, , 
}
for $S_d = 2 \pi^{{1\over 2}d}/\Gamma(\half d)$.
The energy momentum tensor two point function is also of the form given
by \OO\ and can be written
\eqn\TT{\eqalign{
\l & T_{\mu_1\mu_2}(x_1) T_{\nu_1\nu_2} (x_2) \r \cr
&{} = {C_T \over S_d{}^{\! 2}}\, {1\over r_{12}^{\, \, d}} 
\Big ( \half \big ( I_{\mu_1 \nu_1}(x_{12}) I_{\mu_2 \nu_2}(x_{12}) + 
I_{\mu_1 \nu_2}(x_{12}) I_{\mu_2 \nu_1}(x_{12}) \big ) - {1\over d}
\, \de_{\mu_1\mu_2} \de_{\nu_1\nu_2} \Big ) \, ,\cr}
}
with the coefficient $C_T>0$ depending on the particular conformal field
theory.\foot{For $d=2$ with $T(z)=-2\pi T_{zz}(x)$ \ppT\ and \TT\ reduce to the
well known results $\l \phi (x_1) \phi(x_2) T(z_3) \r = \half \De (
z_{12}/z_{13}z_{23})^2 |z_{12}|^{-2\De}$ and $\l T(z_1) T(z_2) \r = 
C_T/(4z_{12}^{\, \, 4})$ so that $\half C_T =c$ the Virasoro central charge.}
The contribution to the four point function is then from \OPEpppp
\eqn\ppppT{
\l \phi(x_1) \phi(x_2) \phi'(x_3) \phi'(x_4) \r \sim  
{1\over r_{12}^{\,\,\De}\, r_{34}^{\,\,\De'}}\,
{\De \De' d^2\over C_T (d-1)^2} u^{{1\over 2}(d-2)} 
G^{(2)}\big (\half(d-2), \half(d-2),d;u,v\big ) \, .
}

Using \Gtwo\ and \Glfour\ we can give complete expressions in two
and four dimensions. If $d=2$ then in \ppppT\ we have
\eqn\Ttwo{
G^{(2)}(0,0,2;u,v) = - 3 \Big ( 1 + {1\over \eta}(1-\half \eta) \ln
(1-\eta) \Big ) + \hbox{conjugate} \, ,
}
while for $d=4$,
\eqn\Ttwo{
u G^{(2)}(1,1,4;u,v) = - 45 \, {x\big ( 1-\half z + {1\over z}(1-z+
{1\over 6}z^2 ) \ln(1-z) \big ) - z\leftrightarrow x  \over z-x } \, .
}

\newsec{AdS/CFT Integrals}

The metric on $AdS_{d+1}$ defines a Weyl equivalence class of metrics
on the boundary $S^d$ while the isometry group $SO(d+1,1)$ becomes the
conformal group on the boundary. This is at the root of the AdS/CFT
correspondence. Much work 
\refs{\Free,\FreeI,\FreeD,\LiuT,\Liu,\San,\Brodie,\Hoff}
has been undertaken investigating the structure
of the conformally covariant correlation functions for boundary points
$x_i$ obtained in terms Feynman graphs for propagators on $AdS_{d+1}$
linking $x_i$ \Witt. For vertices defined by IIB supergravity this is relevant
to the large $N$ limit of $\N=4$ SYM. The two and three point functions
are dictated by conformal invariance while the four point function involves
functions of the two invariants $u,v$, as defined in \defuv, which may
be matched to the operator product expansion.

We show here how the simplest graphs lead to integrals which may be
reduced to the functions $G$ defined earlier in \defG\ and discussed here
and in \one. With the usual metric on $AdS_{d+1}$ and coordinates
$z=(z_0,x), \, x\in {\Bbb R}^d, \, z_0\in {\Bbb R}_+$,
\eqn\met{
\d s^2 = {1\over z_0^{\,2}}\big ( \d z_0^{\,2} + \d x_\mu \d x_\mu \big )\, ,
}
the boundary corresponds to $z_0=0$ together with the point at infinity
$z_0=\infty$. The bulk/boundary propagator is then \Witt
\eqn\Kdel{
K_\Delta(z,x') = {\Gamma(\Delta)\over 2\pi^{{1\over 2}d}
\Gamma(\Delta +1 - \half d)} \, \tK_\Delta(z,x') \, , \qquad
\tK_\Delta(z,x') = \bigg ({z_0 \over z_0^{\,2} + (x-x')^2} 
\bigg )^{\! \Delta}\, .
}
We are initially interested in integrals defining conformal $N$-point
functions of the form, as defined in \FreeD,
\eqn\defD{
D_{\Delta_1 \dots \Delta_N}(x_1,\dots,x_N) = 
{1\over \pi^{{1\over 2}d}} \int \!
\d^{d+1}z{1\over z_0^{\, d+1}} \, \prod_{i=1}^N \tK_{\Delta_i}(z,x_i) \, .
}
Using standard integral representations for 
$(z_0^{\,2} + (x-x_i)^2)^{-\Delta_i}$ the $z$-integration may be 
undertaken giving
\eqn\DN{
D_{\Delta_1 \dots \Delta_N}(x_1,\dots x_N) = 
{\Gamma(\Sigma-\half d)\over 2 \prod_i \Gamma(\Delta_i)}
\int_0^\infty \!\!\!\! {\textstyle \prod_{i=1}^N }
\d \lambda_i \, \lambda_i{}^{\! \Delta_i-1}
\, {1\over \Lambda^{\raise 2pt\hbox{$\scriptstyle \Sigma$}}} \,
e^{- {1\over \Lambda} \sum_{i<j}\lambda_i \lambda_j r_{ij}} \, ,
}
where $\Lambda=\sum_i \lambda_i$ and
\eqn\defSig{
\Sigma=\half \sum_{i=1}^N \Delta_i \, .
}
A crucial observation of Symanzik \Symanzik, recounted in \one,
is that for integrals of the form \DN, subject to \defSig, then 
$\Lambda$ may be modified, without changing the integral, to the form 
$\sum_i \kappa_i\lambda_i$ for any $\kappa_i$, not all zero,
satisfying $\kappa_i\ge 0$. In particular we may choose 
$\Lambda=\lambda_N$ and the integral may then be directly written in
terms of conformally invariant cross ratios like $u,v$. For $N=3$ we have
\eqn\Dthree{
D_{\Delta_1\Delta_2\Delta_3}(x_1,x_2,x_3) =
{\Gamma(\Sigma-\half d)\over 2 \Gamma(\Delta_1)\Gamma(\Delta_2)
\Gamma(\Delta_3)} \, {\Gamma(\Sigma-\Delta_1)\Gamma(\Sigma-\Delta_2)
\Gamma(\Sigma-\Delta_3)\over 
r_{23}{}^{\raise 2pt\hbox{$\scriptstyle \! \Sigma - \Delta_1$}} \,
r_{13}{}^{\raise 2pt\hbox{$\scriptstyle \! \Sigma - \Delta_2$}} \,
r_{12}{}^{\raise 2pt\hbox{$\scriptstyle \! \Sigma - \Delta_3$}} } \, .
}
The result for $N=4$ may also be expressed in term of a function
of the conformal invariants $u,v$ in the form
\eqnn\Dfour
$$\eqalignno{\!\!\!\!\!\!\!
D_{\Delta_1\Delta_2\Delta_3\Delta_4}(x_1,x_2,x_3,x_4)
={}& {\Gamma(\Sigma-\half d)\over 2 \Gamma(\Delta_1)\Gamma(\Delta_2)
\Gamma(\Delta_3)\Gamma(\Delta_4)}\,
{r_{14}{}^{\raise 2pt\hbox{$\scriptstyle \!\! \Sigma-\Delta_1-\Delta_4$}}
r_{34}{}^{\raise 2pt\hbox{$\scriptstyle \!\! \Sigma-\Delta_3-\Delta_4$}}
\over r_{13}{}^{\raise 2pt\hbox{$\scriptstyle \!\! \Sigma-\Delta_4$}}
\, r_{24}{}^{\raise 2pt\hbox{$\scriptstyle \!\! \Delta_2$}}}\cr
&{}\times \oD_{\Delta_1\Delta_2\Delta_3\Delta_4}(u,v) \, . & \Dfour \cr}
$$
Using Symanzik's procedure for evaluating the integrals we have
\eqn\DfourH{
\oD_{\Delta_1\Delta_2\Delta_3\Delta_4}(u,v) =
H\big (\Delta_2,\Sigma-\Delta_4,\Delta_1+\Delta_2-\Sigma+1,
\Delta_1+\Delta_2;u,v\big )\, ,
}
where the function $H$, defined in \one, is directly related
to the function $G$ given by the power series in $u,1-v$ in \defG\ through
\eqn\defH{\eqalign{
H(&\alpha,\beta,\gamma,\delta;u,v) = H(\beta,\alpha,\gamma,\delta;u,v)\cr
&{} = {\Gamma(1-\gamma)\over \Gamma(\delta)}
\Gamma(\alpha) \Gamma(\beta) \Gamma(\delta-\alpha) \Gamma(\delta-\beta) \,
G(\alpha,\beta,\gamma,\delta;u,1-v) \cr
& \ {}+ {\Gamma(\gamma-1)\over \Gamma(\delta-2\gamma+2)}
\Gamma(\alpha-\gamma+1) \Gamma(\beta -\gamma+1) \Gamma(\delta-\gamma-\alpha+1)
\Gamma(\delta-\gamma-\beta+1) \cr
& \qquad\qquad\qquad{}\times u^{1-\gamma}G(\alpha-\gamma+1, \beta -\gamma+1,
2-\gamma, \delta-2\gamma+2;u,1-v) \, . }
}
The symmetry properties of the integral \DN, for $N=4$, are reflected
in various identities obeyed by $H$ which are listed in appendix C.

Besides \Dfour\ we may also consider the scalar exchange contribution to
the four point function which is given by
\eqn\Sfour{\eqalign{\!\!\!\!\!\!\!
S_{\Delta_1\Delta_2\Delta_3\Delta_4}^{\Delta}(x_1,x_2,x_3,x_4)
&{} = {1\over \pi^{{1\over 2}d}} \int \!
\d^{d+1}z{1\over z_0^{\, d+1}} \! \int \!\d^{d+1}w{1\over w_0^{\, d+1}} \,
\tK_{\Delta_1}(z,x_1) \tK_{\Delta_2}(z,x_2) \cr
& \qquad\qquad\qquad\qquad{}\times G_\Delta (z,w) 
\tK_{\Delta_3}(w,x_3) \tK_{\Delta_4}(w,x_4) \, , \cr}
}
for $G_\Delta$ the scalar Green function on $AdS_{d+1}$,
\eqn\defGd{
\big ( - \nab^2 + \Delta (\Delta -d) \big ) G_\Delta (z,w) 
= z_0^{\, d+1} \de^{d+1}(z-w)\, .
}
The explicit form for $G_\Delta$ is unnecessary here except for
\eqn\limG{
G_\Delta (z,w) \sim z_0^{\, \Delta} {1\over 2\pi^{{1\over 2}d}}
{\Gamma(\Delta) \over \Gamma(\Delta-\half d +1)}
\tK_\Delta (w,x) \quad \hbox{as} \quad z_0 \to 0 \, .
}
In consequence
\eqn\defR{
R(z;x_3,x_4) = \int \!\d^{d+1}w{1\over w_0^{\, d+1}}
G_\Delta (z,w) \tK_{\Delta_3}(w,x_3) \tK_{\Delta_4}(w,x_4) \, , 
}
satisfies
\eqn\Rd{
\big ( - \nab^2 + \Delta (\Delta -d) \big ) R(z;x_3,x_4) 
= \tK_{\Delta_3}(z,x_3) \tK_{\Delta_4}(z,x_4) \, ,
}
with the boundary condition, obtained by using \limG\ inside the
integral
\eqn\Rb{
R(z;x_3,x_4) \sim z_0^{\, \Delta} \, 
{\Gamma(\Delta) \over 2\Gamma(\Delta-\half d +1)} \,
D_{\Delta \Delta_3 \Delta_4}(x,x_3,x_4) \quad \hbox{as} \quad z_0 \to 0 \, .
}

To solve \Rd\ with \Rb\ we make use of
\eqn\dKK{\eqalign{
\big ({- \nab^2}&{} + (\Delta_3 + \Delta_4)(\Delta_3 + \Delta_4-d)\big)
\big ( \tK_{\Delta_3}(z,x_3) \tK_{\Delta_4}(z,x_4) \big )\cr
&{}= 4 \Delta_3 \Delta_4 \,\tK_{\Delta_3+1}(z,x_3) \tK_{\Delta_4+1}(z,x_4) 
\, r_{34} \, . \cr}
}
With the aid of \dKK\ we may then write a series solution for $R$ as
\eqnn\solR
$$\eqalignno{\!\!\!\!
R(z;x_3,x_4)&{} = -\quar \sum_{s=0} {(\Delta_3)_s (\Delta_4)_s \over
(\half (\Delta_3 + \Delta_4 - \Delta))_{s+1}
(\half (\Delta_3 + \Delta_4 + \Delta-d))_{s+1}}\cr
&\qquad \qquad \qquad \qquad \qquad \qquad \qquad {}\times
\tK_{\Delta_3+s}(z,x_3) \tK_{\Delta_4+s}(z,x_4)\, r_{34}^{\,\, s}  \cr
{} + {}& \quar \Gamma(\half (\Delta_3 + \Delta_4 + \Delta-d))
{\Gamma(\half(\Delta_3 + \Delta_4 - \Delta))
\Gamma(\half(\Delta+\Delta_{34})) \Gamma(\half(\Delta-\Delta_{34}))
\over \Gamma(\Delta-\half d +1) \Gamma(\Delta_3)\Gamma(\Delta_4)}\cr
\times {}& \sum_{s=0} {(\half(\Delta+\Delta_{34}))_s 
(\half(\Delta-\Delta_{34}))_s \over s! (\Delta-\half d +1)_s }\cr
&\qquad \qquad {}\times
\tK_{{1\over 2}(\Delta+\Delta_{34})+s}(z,x_3)
\tK_{{1\over 2}(\Delta-\Delta_{34})+s}(z,x_4)
\, r_{34}^{\,\, {1\over 2}(\Delta-\Delta_3-\Delta_4)+s}\, . &\solR  \cr}
$$
The series in \solR\ are convergent for sufficiently small $z_0$ or $r_{34}$.
The first term in \solR\ satisfies the inhomogeneous equation \Rd\ while
the second term obeys the corresponding homogeneous equation but
reproduces the required boundary behaviour \Rb\ after using \Dthree. This term
generates the dominant contribution as $z_0 \to 0$ if $\Delta<\Delta_3 + \Delta_4$
which is necessary for the derivation of \Rb\ to be valid.
If $\Delta_3+\Delta_4 - \Delta = 2n, \ n=1,2,\dots$ the two series cancel
except for a finite number of terms and we get
\eqn\soln{
R(z;x_3,x_4)= \quar(n-1)! \! \sum_{s=-n}^{-1} {(\Delta_3)_s (\Delta_4)_s \over
(n+s)! (\Delta - \half d +n)_{s+1}} \,
\tK_{\Delta_3+s}(z,x_3) \tK_{\Delta_4+s}(z,x_4)\, r_{34}^{\,\, s} \, ,
}
which coincides with the solution obtained in \FreeI. Using \solR\ in \Sfour\
we may obtain from \Dfour
\eqn\SS{\eqalign{
S_{\Delta_1\Delta_2\Delta_3\Delta_4}^{\Delta}(x_1,x_2,x_3,x_4) ={}&
{1\over 8\Gamma(\Delta_1) \Gamma(\Delta_2)\Gamma(\Delta_3)\Gamma(\Delta_4)}
{r_{14}{}^{\raise 2pt\hbox{$\scriptstyle \!\! \Sigma-\Delta_1-\Delta_4$}}
r_{34}{}^{\raise 2pt\hbox{$\scriptstyle \!\! \Sigma-\Delta_3-\Delta_4$}}
\over r_{13}{}^{\raise 2pt\hbox{$\scriptstyle \!\! \Sigma-\Delta_4$}}
\, r_{24}{}^{\raise 2pt\hbox{$\scriptstyle \!\! \Delta_2$}}}\cr
&{}\times \S_{\Delta_1\Delta_2\Delta_3\Delta_4}^{\Delta}(u,v) \, ,\cr}
}
where
\eqn\SSs{\eqalign{
\!\!\!\! \S_{\Delta_1\Delta_2\Delta_3\Delta_4}^{\Delta}(u,v) ={}& 
- \sum_{s=0} {\Gamma(\Sigma - \half d +s )\over
(\half (\Delta_3 + \Delta_4 - \Delta))_{s+1}
(\half (\Delta_3 + \Delta_4 + \Delta-d))_{s+1}}\cr
&\quad \ {}\times 
H(\Delta_2, \Sigma - \Delta_4, \Delta_1 + \Delta_2 - \Sigma + 1 -s,
\Delta_1 + \Delta_2 ; u,v) \cr
&{} + {\Gamma(\half (\Delta_3 + \Delta_4 + \Delta-d))
\Gamma(\half(\Delta_3 + \Delta_4 - \Delta)) \over \Gamma(\Delta-\half d +1)}\cr
&{}\times
\sum_{s=0} { \Gamma(\half (\Delta_1 + \Delta_2+ \Delta-d) + s) \over
s! \, (\Delta-\half d +1)_s} \cr
&\quad \ {}\times
H(\Delta_2, \Sigma - \Delta_4, \half (\Delta_1 + \Delta_2 - \Delta) + 1 -s,
\Delta_1 + \Delta_2 ; u,v) \, .  \cr}
} 
As in \soln\ it is easy to verify that this reduces to a finite sum
if $\Delta_3+\Delta_4 - \Delta = 2n$. For general $\Delta$ we may use
\defH\ and \defG\ to rewrite \SSs\ as the sum of three terms with
different leading powers in $u$ \HoffR,
\eqnn\Sthree
$$\eqalignno{
& u^{{1\over 2}(\Delta_1+\Delta_2)} 
\S_{\Delta_1\Delta_2\Delta_3\Delta_4}^{\Delta}(u,v) \cr
&{} =\Gamma(\half(\Delta_1+\Delta_2-\Delta))
\Gamma(\half(\Delta_3+\Delta_4-\Delta))\Gamma(\half(\Delta_1+\Delta_2+\Delta-d))
\Gamma(\half(\Delta_3+\Delta_4+\Delta-d))\cr
&\quad {}\times {1\over \Gamma(\Delta)\Gamma(\Delta+1-\half d)}
\Gamma(\half(\Delta+\Delta_{12}))\Gamma(\half(\Delta-\Delta_{12}))
\Gamma(\half(\Delta+\Delta_{34}))\Gamma(\half(\Delta-\Delta_{34})) \cr
&\quad {}\times u^{{1\over 2}\Delta}
G\big (\half\Delta-\half\Delta_{12},\half\Delta+\half\Delta_{34},
\Delta+1-\half d, \Delta;u,1-v\big ) \cr
&\qquad {} - u^{{1\over 2}(\Delta_3+\Delta_4)}\,
\G_{\Delta_1\Delta_2\Delta_3\Delta_4}^{\Delta}(u,v)
- u^{{1\over 2}(\Delta_1+\Delta_2)}
\G_{\Delta_4\Delta_3\Delta_2\Delta_1}^{\Delta}(u,v) \, , & \Sthree \cr}
$$
for
\eqnn\defGG
$$\eqalignno{
\G_{\Delta_1\Delta_2\Delta_3\Delta_4}^{\Delta}(u,v)= {}&
{\Gamma(\Sigma-\half d)\over \Gamma(\Delta_3+\Delta_4)}
\Gamma(\Sigma-\Delta_3-\Delta_4)\Gamma(\Delta_3)\Gamma(\Delta_4)
\Gamma(\Sigma-\Delta_1)\Gamma(\Sigma-\Delta_2) \cr
\times {}& \sum_{m,n=0} {(\Sigma-\Delta_2)_m (\Delta_4)_m G^\Delta_m \over m!
(\Sigma-\Delta_1-\Delta_2+1)_m} \,
{(\Delta_3)_{m+n} (\Sigma-\Delta_1)_{m+n}\over (\Delta_3+\Delta_4)_{2m+n}}
\, u^m(1-v)^n \, , \cr
G^\Delta_m = \sum_{s=0}^m (-1)^s & {m!\over (m-s)!} \,
{(\Sigma - \half d)_s \over (\half(\Delta_3+\Delta_4-\Delta))_{s+1}
(\half(\Delta_3+\Delta_4+\Delta- d))_{s+1} } \, . & \defGG \cr}
$$
To achieve the desired form for 
$\G_{\Delta_4\Delta_3\Delta_2\Delta_1}^{\Delta}$ requires non
trivial relations for ${}_3F_2$ functions with argument 1.
The first term in \Sthree\ matches exactly the contribution of
a scalar operator with dimension $\Delta$ in the operator product
expansion. From \defGG\ $\G_{\Delta_1\Delta_2\Delta_3\Delta_4}^{\Delta}(u,v)
=v^{-\Delta_3}\G_{\Delta_2\Delta_1\Delta_3\Delta_4}^{\Delta}(u/v,1/v)
=v^{-\Sigma+\Delta_1}\G_{\Delta_1\Delta_2\Delta_4\Delta_3}^{\Delta}(u/v,1/v)$.
Furthermore $u^{{1\over 2}(\Delta_1+\Delta_2)} 
\S_{\Delta_1\Delta_2\Delta_3\Delta_4}^{\Delta}(u,v) =
u^{{1\over 2}(\Delta_3+\Delta_4)} 
\S_{\Delta_4\Delta_4\Delta_2\Delta_1}^{\Delta}(u,v) $ as well as satisfying
relations for $\Delta_1\leftrightarrow \Delta_2$ or
$\Delta_3\leftrightarrow \Delta_4$ and $u\to u/v, \, v\to 1/v$ which
follow from the preceding relations for 
$\G_{\Delta_1\Delta_2\Delta_3\Delta_4}^{\Delta}$.
The result given by \Sthree\ and \defGG\ provides a
representation valid for $u\sim 0, \, v\sim 1$ and corresponds essentially
with that given by Liu \Liu. In principle \SSs\ may be used to find
a form appropriate for $u\sim 1, \, v\sim 0$ but, for the general case,
the results are significantly more complicated \Hoff.
The last line in \Sthree\ corresponds to the contribution of
operators with dimensions $\Delta_3+\Delta_4+ 2n, \, \Delta_1+\Delta_2+ 2n, \
n=0,1,2,\dots$.

\newsec{Results for Free Fields}

With the explicit formula \Glfour\ for the contribution for arbitrary
spin operators to the four point function in four dimensions then
it is natural to consider 
the associated partial wave expansion for some simple conformally invariant
expressions for the four point function. We consider here some examples
arising in free field theories
and verify consistency with the expected form of the operator
product expansion.

For the general case considered here we relax the normalisation assumption
of section 2 and consider a quasi-primary scalar field $\phi$, with
scale dimension $\Delta_\phi$, for which the two point function has the
form
\eqn\phitwo{
\l \phi(x_1) \phi(x_2) \r = {N_\phi\over r_{12}^{\,\, \Delta_\phi}} \, .
}
The corresponding four point function is then taken to be
\eqn\phifour{
\l \phi(x_1) \phi(x_2) \phi(x_3) \phi(x_4) \r = N_\phi{}^{\! 2} \,
{1\over r_{12}^{\,\, \Delta_\phi}\, r_{34}^{\,\, \Delta_\phi}} 
\big ( 1 + g_\phi(u,v) \big ) \, ,
}
where $g_\phi(u,v)$ satisfies the symmetry conditions
\eqn\gsym{
g_\phi(u,v) = g_\phi(u/v,1/v) \, , \qquad 1 + g_\phi(u,v) =
\Big ( {u\over v} \Big )^{\! \Delta_\phi} \big ( 1 + g_\phi(v,u)\big ) \, ,
}
and we may assume $g_\phi(0,v)=0$.
The $1$ in \phifour\ is then the leading singularity as $r_{12}\to 0$
and represents just the contribution of the identity 
operator in the operator product expansion of $\phi(x_1) \phi(x_2)$. 
By virtue of \OPEpppp\ $g_\phi(u,v)$ should be expanded as
\eqn\OPEa{
g_\phi(u,v) = \sum_{\Delta,\ell} c_{\Delta,\ell} \,
u^{{1\over 2}(\Delta-\ell)}
G^{(\ell)}\big (\half(\Delta - \ell),\half(\Delta - \ell),\Delta;u,v\big )\,,
}
with $\Delta>0$. The set of $\Delta,\ell$ which are necessary in
\OPEa\ determines the spectrum of operators which contribute to the
operator product expansion of $\phi(x_1) \phi(x_2)$.
As a consequence of \symG\ and the first relation in
\gsym\ $\ell$ must be even.

For free field theories $g_\phi(u,v)$ is analytic in $u,1-v$ which
corresponds to requiring that any $\Delta$ contributing to the sum
in \OPEa\ is an even integer as well. To apply the explicit result
\Glfour\ we first define, with the definitions \uvxz,
\eqn\defhp{
h_\phi(z,x) = {z-x \over u} g_\phi(u,v) \, ,
}
where
\eqn\hhs{
h_\phi(z,x) = - h_\phi(x,z) = - h_\phi(z',x')  \, , \qquad
x' = {x\over x-1} \, , \quad z' = {z\over z-1} \, .
}
Writing $\ell =2m, \ \Delta - \ell = 2(t+1)$, then, with \Glfour,
\OPEa\ becomes
\eqn\OPEb{
h_\phi(z,x) = \sum_{t=0} \Big ( H_t(z) \, x^t F(t,t;2t; x) -
H_t(x) \, z^t F(t,t;2t; z) \Big ) \, ,
}
where, with $c_{\Delta,\ell}\to c_{mt}$ for $m,t$ integers, 
\eqn\defHt{
H_t(z) = \sum_{m=0} c_{mt} \, {1\over 2^{2m}} z^{2m+t+1} F(2m+t+1, 2m+t+1;
4m+2t+2;z) \, .
}
By virtue of standard hypergeometric identities this satisfies
\eqn\symH{
H_t(z) = (-1)^{t+1} H_t (z') \, .
}
Furthermore if we compare with \ppppT, with $\De=\De'=\De_\phi$, $d=4$,
the contribution of the energy momentum tensor with $\ell=2$ corresponds
to
\eqn\cen{
c_{10} = {16 \De_\phi{}^{\! 2} \over 9 C_{T}} \, .
}

To obtain an algorithm for determining $c_{mt}$ it is convenient to
write firstly
\eqn\hhh{
h_\phi(z,x) = h_0(z) - h_0(x) + \hh_\phi(z,x) \, ,
}
where $\hh_\phi(z,x)$ is ${\rm O}(zx)$ and $h_0(z) = - h_0(z')$, from \hhs.
We also expand  $\hh_\phi(z,x)$ in powers of $x$, 
\eqn\exh{
\hh_\phi(z,x) = \sum_{n=1} \hh_n(z) x^n \, .
}
The essential equation \OPEb\ may then be decomposed into independent equations
for each power in $x$. For the terms of ${\rm O}(x^0)$ we get
\eqn\eqcz{
h_0(z) = H_0(z) \, ,
}
which determines $c_{m0}$. 
For the terms  ${\rm O}(x^n)$, $n>0$, we have 
\eqn\eqct{\eqalign{
& \hh_n(z) - \sum_{t=1}^{n-1} {\big ( (t)_{n-t} \big )^2 \over (n-t)!
(2t)_{n-t} } H_t(z)\cr 
&{} +\sum_{t=1}^{n-1} \sum_{m=0}^{[{1\over 2}(n-t-1)]}\!\!\! c_{mt} \,
{1\over 2^{2m}} {\big ( (2m+t+1)_{n-2m-t-1} \big )^2 \over (n-2m-t-1)!
(4m+2t+1)_{n-2m-t-1} } \, z^t F(t,t;2t; z) = H_n(z) \, , \cr}
}
which expresses $c_{mn}$ recursively in terms of $c_{mn'}, \ n'<n$. For $n=1$
\eqct\ becomes just $\hh_1(z) = H_1(z)$ determining $c_{m1}$. 
With further manipulation this can be written as
\eqn\eqctn{
\hh^{(n)}(z) + \sum_{m=0}^{[{1\over 2}n]-1} c_{mt}{1\over 2^{2m}}\, 
z^t F(t,t;2t; z)\bigg |_{t=n-1-2m} = H_n(z) \, ,
}
where, as also in \eqct, $[{1\over 2}n]$ denotes the integer part and
\eqn\eqhn{
\hh^{(n)}(z) = \hh_n(z) - \sum_{t=1}^{n-1} 
{\big ( (t)_{n-t} \big )^2 \over (n-t)! (2t)_{n-t} } \hh^{(t)}(z) 
= \sum_{r=0}^{n-1} (-1)^r
{\big ( (n-r)_{r} \big )^2 \over r! (2n-r-1)_{r} } \hh_{n-r}(z) \, .
}
It is crucial that the left hand side of \eqctn\ is compatible with \symH.
For the terms involving $z^t F(t,t;2t; z)$ this is automatic with the
restriction $t=n-1-2m$.  The condition $\hh^{(n)}(z)=(-1)^{n+1}\hh^{(n)}(z')$
follows from \eqhn\ together with 
\hhs\ which, with \hhh\ and \exh, implies $\hh_n(z) = - \sum_{r=1}^{n} (-1)^r
(r)_{n-r} \hh_r(z')/(n-r)!$. It is also necessary, although less evident,
that the left hand side of \eqctn\ is ${\rm O}(z^{n+1})$ to match the
leading term on the right hand side.

The simplest case is that for a free scalar field,  $\phi \to \vphi$ satisfying
$\pr^2 \vphi =0$ and $\Delta_\vphi=1$. 
With canonical normalisation in \phitwo\ we have $N_\vphi = 1/4\pi^2$. For 
this case it is easy to calculate that in \phifour
\eqn\vgp{
g_\vphi(u,v) = u + {u\over v} \, ,
}
so that in \hhh
\eqn\hphi{
h_0(z) = z + {z\over 1-z} \, , \qquad \hh_\vphi(z,x) =0 \, .
}
The only equation to solve is then \eqcz.
By using algebraic manipulation programmes we have verified that
\eqn\solc{
c_{m0} = 2^{2m+1} {\big ( (2m)! \big )^2 \over (4m)!} \, ,
}
is compatible with direct calculations
for the first 20 terms. Although a formal proof of \solc\ is doubtless
possible we have not invested the effort necessary to achieve it.
Reassuringly $c_{m0}>0$ as required by unitarity. Furthermore 
$c_{10}={4\over 3}$ and then \cen\ gives $C_{T} = {4\over 3}$
which is the correct value for the free scalar theory in four 
dimensions \hughone. Since $c_{mt}=0, \, t\ge1$ only operators with twist
$(\Delta - \ell)$ two are present in the operator product expansion. This is 
also as expected since they are just 
$\vphi \pr_{\mu_1} \dots \pr_{\mu_\ell} \vphi$ in free field theory.

A further example in free scalar theory arises for $\phi \to \half \vphi^2$,
where we have $\Delta_{{1\over 2}\vphi^2}=2$ and in \phitwo\
$N_{{1\over 2}\vphi^2}= 1/32\pi^4$. In the corresponding four point
expression \phifour\ we then have
\eqn\mfour{
g_{{1\over 2}\vphi^2}(u,v) = u^2 + {u^2\over v^2} + C \Big (
u + {u\over v} + {u^2\over v} \Big )  = g_C(u,v) + C \Big (
u + {u\over v} \Big ) \, ,
}
for
\eqn\gC{
g_C(u,v) = u^2 + {u^2\over v^2} + C \, {u^2\over v} \, .
}
For a single scalar field $C=4$, more generally for $N_s$ free scalar fields
$C=4/N_s$. 
It is convenient then to allow for arbitrary values of the paramter $C$.
The terms which are ${\rm O}(u)$ give in this case
\eqn\hpp{
h_0(z) = C \Big ( z + {z\over 1-z} \Big ) \, .
}
Applying \defhp\ and \hhh\ to $g_C(u,v)$ as given in
\gC\ gives
\eqn\hmf{
\hh_{C}(z,x) =  
(z-x)zx \Big ( 1+ {1\over (1-z)^2(1-x)^2} + {C\over (1-z)(1-x)} \Big )\,. 
}
The expansion in powers of $x$ is straightforward and from \eqhn\ we
find
\eqn\fnn{
\hh^{(n)}_C(z) = \cases{{\displaystyle{(n-1)!^2\over (2n-2)!} \,
\Big ( z^2 + z'^2 + \big (n(n-1)-C\big) (z+z') \Big )} \, , & $n$ odd, \cr
- {\displaystyle{(n-1)!^2\over (2n-2)!} \, 
\Big ( z^2-z'^2 + \big ( n(n-1)+C \big )(z-z')\Big ) } \, , & $n$ even, \cr}
}
for $z'=z/(z-1)$. The required symmetry under $z \leftrightarrow z'$ is
evident. The result for $c_{m0}$ obtained from \eqcz\ is clearly $C$ 
times that given by \solc, which if $C=4$ gives the correct value of 
$c_{10}$ for the same value of $C_T$. A similar approach for solving
\eqctn\ to that leading to \solc\ suggests, for any $C$,
\eqn\cmt{
c_{mt} = 2^{2m} {(2m+t-1)!\, (2m+t)!\, \big((t-1)!\big )^2\over
(4m+2t-1)! \, (2t-2)!} \big (2(2m+1)(m+t) + (-1)^{t+1}C \big ) \, , \quad
t\ge 1 \, .
}
This is positive for $-2<C<4$. 
If $C=4$ then $c_{02}=0$ which reflects the fact that
the only potential operator with $\ell=0, \, \Delta=6$, $\vphi^2\pr \vphi
{\cdot \pr \vphi}$ is a descendant of $\vphi^4$.

For the case of a free fermion field $\psi$ we may consider
$\phi \to \bpsi \psi$, which is a scalar operator with $\De_{\bpsi \psi}=3$.
the basic two point function is
\eqn\twopsi{
\l \psi(x_1) \bpsi(x_2) \r = {\gamma\, {\cdot x_{12}} \over 2\pi^2 \, 
r_{12}^{\, \, 2} } \, ,
}
from which in \phitwo\ $N_{\bpsi \psi} = 1/\pi^4$. In \phifour\
\eqn\fourpsi{
g_{\bpsi \psi}(u,v) = u^3 + {u^3\over v^3} + {1\over 4}\Big ( u(v-1-u)
+{u\over v^2} (1-u-v) + {u^3\over v^2}(u-1-v) \Big ) \, .
}
{}From \defhp\ and \hhh,
\eqn\hpsi{
h_0(z)={1\over 4} \, {z^3\over (1-z)^2}(2-z) \, ,
}
and
\eqn\hhpsi{
\hh_{\bpsi \psi}(z,x)= (z-x)z^2x^2 \bigg ( 1 + {1\over (1-z)^3(1-x)^3}
-{1\over 4(1-z)^2(1-x)} - {1\over 4(1-z)(1-x)^2} \bigg ) \, .
}
{}From \exh\ and \eqhn\ in this case
\eqn\fnnF{
\hh^{(n)}(z) = \cases{-{\displaystyle{(n-1)!n!\over 8(2n-3)!}
\Big ( 4(z^3+z'^3) + \big (n(n-1) -3\big )(z^2 + z'^2) \Big )} \, , 
& $n$ odd, \cr
{\displaystyle{(n-1)!n!\over 8(2n-3)!} \, 
\Big ( 4(z^3-z'^3) + \big (n(n-1)- 1 \big )(z^2-z'^2)\Big )}\, ,& $n$ even. \cr}
}
In a similar fashion as previously we have found for $t=0$
\eqn\cmtf{
c_{00}=0 \, , \qquad
c_{m0} = 2^{2m-2}{(2m)!(2m+1)!\over (4m-1)!} \, , \quad m=1,2,\dots \, ,
}
and for $t=1,2,\dots$,
\eqn\cmtff{
c_{mt} =   2^{2m-3} {(2m+t+1)!\, (2m+t)!\, (t-1)! \, t! \over
(4m+2t-1)! \, (2t-3)!} \big ( 2(2m+1)(m+t) + (-1)^{t+1} \big ) \, .
}
The operators for $t=0$ are of the form $\bpsi \gamma_{(\mu_1} \pr_{\mu_2}
\dots \pr_{\mu_\ell )}\psi$ and the leading operator which contributes is
the $\ell=2$ energy momentum tensor. In this theory $c_{10} = 2$ so that
\cen\ gives $C_T=8$, as expected for free Dirac fermions. From \cmtff\
we may note that $c_{m1}=0$ which follows from \hhpsi\ since the leading
term is ${\rm O}(z^2x^2)$. This may be explained since the relevant
operators are $\bpsi \pr^r \psi \, \bpsi \pr^s \psi $ with $\Delta = 6+r+s$
and $r+s\ge \ell$ so that the minimal $t$ for operators of this form
is $t=2$.

We may also consider the case of free vector fields when we may take
$\phi \to {1\over 4} F^2 = {1\over 4} F_{\mu\nu}F_{\mu\nu}$, the lowest
dimension scalar operator with $\De_{{1\over 4} F^2}=4$. The essential
two point function is
\eqn\twoF{
\l F_{\mu\nu}(x_1) F_{\sigma \rho}(x_2) \r = {1\over \pi^2\, r_{12}^{\, \, 2}}
\big ( I_{\mu\sigma}(x_{12}) I_{\nu\rho}(x_{12}) - I_{\mu\rho}(x_{12}) 
I_{\nu\sigma}(x_{12})  \big ) \, ,
}
where the inversion tensor is given by \defI. It is easy to find in
\phitwo\ $N_{{1\over 4} F^2}= 3/\pi^4$. In this case we may calculate
in \phifour\ (the derivation is sketched in appendix D),
\eqn\fourF{
g_{{1\over 4} F^2}(u,v) = u^4 + {u^4\over v^4}
+ {2\over 9} \Big ( u(v-1-u)^2
+{u\over v^3} (1-u-v)^2 + {u^4\over v^3}(u-1-v)^2 - u^2 - {u^2\over v^2}
- {u^4\over v^2} \Big ) \, .
}
{}From \defhp\ and \hhh,
\eqn\hFF{
h_0(z) = {2\over 9} \bigg ( z^3 + {z^3\over (1-z)^3} \bigg ) \, .
}
and
\eqn\hhFF{\eqalign{
\hh_{{1\over 4} F^2}(z,x)= {}& 
(z-x)z^3x^3 \bigg ( 1 + {1\over (1-z)^4(1-x)^4}\cr
&{} +{2\over 9}\Big ({1\over (1-z)(1-x)^3}+{1\over (1-z)^2(1-x)^2}
+{1\over (1-z)^3(1-x)}\Big ) \bigg ) \, . \cr}
}
As previously we may calculate
\eqn\fnFF{
\hh^{(n)}(z) = \cases{(n-2){\displaystyle{(n-1)!(n+1)!\over 72(2n-3)!}
\Big ( 9(z^4+z'^4) + \big (n(n-1) - 8 \big )(z^3 + z'^3) \Big )} \, ,
& $n$ odd, \cr
-(n-2){\displaystyle{(n-1)!(n+1)!\over 72(2n-3)!} \, 
\Big ( 9(z^4-z'^4) + \big (n(n-1)- 4 \big )(z^3-z'^3) \Big ) } \, , & 
$n$ even. \cr}
}
Just as before we then find
\eqn\cmtF{
c_{00}=0 \, , \qquad
c_{m0} = 2^{2m-1}(2m-1){(2m)!(2m+2)!\over 9(4m-1)!}\, , \quad m=1,2,\dots \, ,
}
and for $t=1,2,\dots$,
\eqn\cmtFF{\eqalign{
c_{mt} =  (t-2) & 2^{2m-4} {(2m+t+2)!\, (2m+t)!\, (t-1)! \, (t+1)!\over
9(4m+2t-1)! \, (2t-3)!} (2m+t-1) \cr
&{} \times \big ( (2m+1)(m+t) + (-1)^{t+1} \big ) \, . \cr}
}
For $t=0$ the relevant operators are $F_{(\mu_1|\nu|} \pr_{\mu_2} \dots
\pr_{\mu_{\ell-1}}F_{\mu_\ell)\nu}$, with traces subtracted. For $\ell=2$,
giving the energy momentum tensor,
we have $c_{10}=16/9$ so that \cen\ gives $C_T=16$, in accord with
the required result for free vector theories. It should also be noted
that the operator $\quar F^2$ does not appear in the operator product
expansion of $\quar F^2(x_1) \quar F^2(x_2)$. We also have in
\cmtFF\ $c_{m1}=c_{m2}=0$ which has a similar explanation as for the fermion
case, the $F^4$ operators which are present have $\Delta \ge \ell +8$.

Finally we consider an example based of a four point function formed by
scalar operators $T^{ij}=T^{ji}$ and its conjugate $\bT_{ij}$ which, with
$i,j=1,2$ $SU(2)_R$ indices, have dimensions $\Delta_T=\Delta_\bT=2$ and
are the lowest dimension scalar operators in a $\N=2$ supersymmetric theory 
\Eden. The relevant four point function has the general form
\eqnn\fourT
$$\eqalignno{
\l T^{i_1 j_2}(x_1) &\bT_{i_2 j_2} (x_2) T^{i_3 j_3}(x_3) 
\bT_{i_4 j_4} (x_4) \r \cr
= {}& \de^{(i_1}{}_{i_2} \de^{j_1)}{}_{j_2} \, \de^{(i_3}{}_{i_4} 
\de^{j_3)}{}_{j_4} \, {1\over r_{12}^{\, \, 2}\,  r_{34}^{\, \, 2} } \, a(u,v)
+ \de^{(i_1}{}_{i_4} \de^{j_1)}{}_{j_4} \, \de^{(i_3}{}_{i_2} \de^{j_3)}{}_{j_2}
\, {1\over r_{14}^{\, \, 2}\,  r_{23}^{\, \, 2} } \, b(u,v) \cr
&{}+ \de^{(i_1}{}_{(i_2} \de^{j_1)}{}_{(i_4} \, \de^{(i_3}{}_{j_4)} 
\de^{j_3)}{}_{j_2)} \, 
{1\over r_{12}\, r_{23}\, r_{34}\, r_{14} } \, c(u,v) \, , & \fourT \cr}
$$
with $a(u,v)=b(v,u), \ c(u,v)=c(v,u)$. In the free case $a,b,c$ are
constants and setting, with suitable normalisation $a=b=1$, the conformal
invariant four point functions for $T\bT$ projected on $R=0,1,2$
representations are
\eqn\AR{
A_0(u,v) = 1 + {\thir}\, {u^2\over v^2} + \half c \, {u\over v} \, , \quad
A_1(u,v) = {u^2\over v^2} + {u\over v} \, , \quad A_2(u,v) = {u^2\over v^2}
\, .
}
For this case there is no symmetry under $u\leftrightarrow u/v, \, v
\leftrightarrow 1/v$ but it is straightforward to decompose each term
arising in \AR\ into even and odd pieces. For the even pieces, $u^2/v^2
+ u^2$ and $u/v+u$, the relevant expansion coefficients are given
by \cmt, for $C=0$, and \solc.\foot{Perhaps we may note that for
$(u/v)^n +u^n$ the expansion coefficients $c_{mt}^{(n)}$ are given by
$$
c_{mt}^{(n)} = 2(2m+1)(m+t) \Big ({ 2^m (t-1)!\over (n-1)!(n-2)!}\Big )^{\! 2}
{(2m+t-1)!\, (2m+t)!\, (2m+t+n-2)! \, (t+n-3)! \over
(4m+2t-1)! \, (2m+t-n+2)! \, (2t-2)! \, (t-n+1)!}  \, .
$$
This is zero for $t=0,1,\dots n-2$, $n\ge 2$. This formula is also correct 
for $n=1$ when only $t=0$ contributes.} For the odd pieces only odd values
of $\ell$ contribute and, with the same definition of $t$ as previously we 
can then write
\eqn\odduv{
{u^2\over v^2} - u^2 = - \sum_{m,t} d_{mt} \, u^{t+1}
G^{(2m+1)}(t+1,t+1,2t+2m+3;u,v) \, ,
}
where the negative sign is a consequence of the fact that the operators
occurring in the operator product expansion are anti-hermitian for
odd $\ell$ in this example. As before we determine
\eqn\dtm{
d_{mt} =   2^{2m+2} {(2m+t+1)!\, (2m+t)!\, \big ( (t-1)! \big )^2 \over
(4m+2t+1)! \, (2t-2)! } (2m+2t+1)(m+1) \, .
}
A similar equation to \odduv\ may be written for $u/v-v$ but in this
case only $t=0$ contributues. This equation may be reduced to
\eqn\dz{
{z\over 1-z} -z = \sum_{m=0} d_m \, {1\over 2^{2m+1}}z^{2m+2}
F(2m+2,2m+2;4m+4;z) \, ,
}
which determines
\eqn\sold{
d_m = 2^{2m+2} {\big ( (2m+1)!\big )^2 \over (4m+2)!} \, .
}

\newsec{Conclusion}

A crucial result of this paper is that it is possible to fine a simple closed
form expression for the contribution of an arbitrary spin operator to
the four point function. For simplicity, taking $\De_1=\De_2, \, \De_3=\De_4$,
the result from \OPEpppp\ and \Glfour\ is
\eqn\OPEs{\eqalign{
& u^{{1\over 2}(\Delta-\ell)} G^{(\ell)}\big (\half(\Delta - \ell),
\half(\Delta - \ell),\Delta;u,v\big ) \cr
&{}= {(zx)^{{1\over 2}(\Delta-\ell)}\over z-x} \Big ((-\half z)^\ell
z F\big (\half(\De+\ell),\half(\De+\ell);\De+\ell;z \big )  \cr
&\qquad \qquad \qquad
{}\times F\big (\half(\De-\ell-2),\half(\De-\ell-2);\De-\ell-2;x\big )
- z \leftrightarrow x \Big ) \, , \cr }
}
for $u=zx, \, v=(1-z)(1-x)$, as in \uvxz.
In the previous section we have shown how this may be applied to identify the
relevant operators in some simple cases based on free field theories. The
variables $z,x$ play an essential role in the expression \OPEs\ and it
would be desirable to find a more direct justification of this result in
which the significance of such a parameterisation of $u,v$ was perhaps more
transparent. Of course individual contributions of the form \OPEs\ do
not have the required form for $v\sim 0, \, u \sim 1$ and this
constrains the contributions of different operators although, as yet,
there is no organising principle as in two dimensions. The critical unitarity
constraint that the invariant function of $u,v$ which describes the four
point function in conformal field theories should be expandible in terms of
contributions of the form \OPEs, for suitable $\Delta$ and $\ell$ and with
positive coefficients in appropriate cases, should become easier to analyse
with the explicit expressions for $G^{(\ell)}$ obtained here and exhibited
in \OPEs. These results should allow further extension of the
analysis of four point functions obtained through the AdS/CFT correspondence
in terms of the operator product expansion in the large $N$ limit
of $\N=4$ supersymmetric gauge theories \refs{\Hok,\OPEC}.

To see the simplifications obtained by using the variables $z,x$ in another 
context we mention also some recent results \Eden\
found through the use of superconformal Ward identities, using the harmonic
superspace formalism, for the four point function exhibited in \fourT.
Although $T^{ij}, \bT_{ij}$ are scalar fields which are the lowest
components of $\N=2$ hypermultiplets the superconformal Ward identity  lead
to constraints  on the dependance of $a,b,c$ on the invariants $u,v$,
\eqn\abcS{\eqalign{
{\pr \over \pr u} c = {}& {v\over u}{\pr\over \pr v}a - {\pr\over \pr v}b
- \bigg ( 1 -{1\over v}+{u\over v}\bigg ) {\pr\over \pr u} b \, , \cr
{\pr \over \pr v} c = {}& {u\over v}{\pr\over \pr u}b - {\pr\over \pr u}a
- \bigg ( 1 -{1\over u}+{v\over u}\bigg ) {\pr\over \pr v} a \, . \cr}
}
Rewriting these equations in terms of $z,x$ gives
\eqn\abczx{
{\pr\over \pr x}c = {1-z\over z} \, {\pr\over \pr x}a + {z\over 1-z} \,
{\pr\over \pr x} b \, , \qquad 
{\pr\over \pr z}c = {1-x\over x} \, {\pr\over \pr z}a + {x\over 1-x} \,
{\pr\over \pr z} b \, .
}
The solutions are then clearly
\eqn\solabc{
c - {1-z\over z} \, a - {z\over 1-z} \, b = f(z) \, , \qquad
c - {1-x\over x} \, a - {x\over 1-x} \, b = f(x) \, ,
}
where we have imposed the essential symmetry under $z\leftrightarrow x$.
Eliminating $c$ or $a$ gives
\eqn\sab{
{1\over u}\, a  - {1\over v}\, b = {f(z)-f(x)\over z-x} \, , \qquad
{1\over v}\, c  - {1-v-u\over v^2}\, b = 
{{z\over 1-z}f(z)-{x\over 1-x}f(x)\over z-x} \, ,
}
which are equivalent to the solutions found in \Eden. To satisfy
the symmetry under $u\leftrightarrow v$ we must have $f(z)=f(1-z)$.
For the free case discussed in section 6 $f(z)=2+c-1/z(1-z)$. Eden
et al \Eden\ have argued that there are no higher order corrections
in the interacting theory. It would be interesting to see the implications
in the context of the operator product expansion.

\bigskip
\noindent{\bf Acknowledgements}

One of us (FAD) would like to thank the EPSRC, the National University of
Ireland and Trinity College, Cambridge for support. He is also very
grateful to David Grellscheid for assistance with Mathematica. HO would
like to thank Anastasios Petkou for email correspondence and very useful
comments.

\vfill\eject
\appendix{A}{Differential Operators for the Operator Product Expansion}

We describe here how to construct differential operators satisfying
\defC\ for $\ell=1,2$. These cases have been considered previously for
$a=b$ and $S=d-1,d$ respectively in \refs{\LW,\OPEC} and results for
any $\ell$ were given in \Ftwo.

For $\ell=1$ and $S=a+b+1$ we use the definition of $Z$ in \defZ\ to write
\eqn\Irep{\eqalign{
{1\over r_{13}^{\, \, a} \, r_{23}^{\, \, b}}\, Z_{\nu}
= {}& {1\over S} {1\over B(a+1,b+1)} \int_0^1 \!\! \d \alpha\,
{\alpha^{a-1}(1-\alpha)^{b-1} \over \big ( \alpha r_{13} +
(1-\alpha) r_{23} \big )^S} \big ( b\alpha \, x_{13} - a (1-\alpha)\,
x_{23} \big ) {}_\nu \cr
= {}& {1\over S} {1\over B(a+1,b+1)} \int_0^1 \!\! \d \alpha\,
\alpha^{a-1}(1-\alpha)^{b-1} \sum_{n=0} {1\over n!}
\big ( A\, \quar \pr_{x_2}^{\,\, 2} \big )^n \cr
& {}\times {1\over y^{2S}}
\bigg ( {S-1 \over (S + 1 -\half d)_n} \alpha(1-\alpha) x_{12\nu}
+ {1\over (S -\half d)_n} \big ( b\alpha - a (1-\alpha)\big )
y_\nu \bigg ) \, , \cr}
}
where
\eqn\defAy{
y = x_{23} + \alpha x_{12} \, , \qquad A=-\alpha(1-\alpha) r_{12} \, ,
}
so that $\alpha r_{13} + (1-\alpha) r_{23} = y^2 +A$, and we have
used, for $\ell=0,1$,
\eqn\diff{
\big (\quar \pr^2 \big)^n {1\over y^{2S}} y_{\nu_1} \dots y_{\nu_\ell}
\C_{\nu_1 \dots \nu_\ell} = (S)_n (S+ 1 -\half d - \ell)_n \,
{1\over y^{2(S+n)}} y_{\nu_1} \dots y_{\nu_\ell}
\C_{\nu_1 \dots \nu_\ell} \, .
}
We now employ
\eqn\SSr{
{S-1\over S} {1\over y^{2S}} x_{12\nu} = {1\over y^{2S}}  x_{12\mu}
I_{\mu\nu}(y) - {1\over S} {\d \over \d \alpha} {1\over y^{2S}} y_\nu \, ,
}
and, after integrating by parts, the relation
\eqn\Srel{\eqalign{
& {1 \over (S + 1 -\half d)_n}\big ( (a+n)(1-\alpha)-(b+n)\alpha \big )
+ {1\over (S -\half d)_n} \big ( b\alpha - a (1-\alpha)\big ) \cr
&\qquad {} = - n {1\over (S -\half d)_{n+1}} 
\big ( (a+1-\half d) \alpha - (b+1-\half d)(1-\alpha) \big ) \, , \cr}
}
to obtain
\eqnn\Ione
$$\eqalignno{
{1\over r_{13}^{\, \, a} \, r_{23}^{\, \, b}}\, Z_{\nu}
= {}& {1\over B(a+1,b+1)} \int_0^1 \!\! \d \alpha\,
\alpha^{a}(1-\alpha)^{b}\cr
& \qquad \qquad \qquad {} \times \sum_{n=0} {1\over n!}
{1 \over (S + 1 -\half d)_n}\,
\big ( A\, \quar \pr_{x_2}^{\,\, 2} \big )^n 
{1\over y^{2S}} x_{12\mu} I_{\mu\nu}(y) \cr
{}+{}& {r_{12} \over B(a+1,b+1)} \int_0^1 \!\! \d \alpha\,
\alpha^{a}(1-\alpha)^{b}
\big ( (a+1-\half d) \alpha - (b+1-\half d)(1-\alpha) \big ) \cr
& \qquad \qquad \qquad {} \times \sum_{n=0} {1\over n!}
{1 \over (S + 1 -\half d)_{n+1}}\, \big ( A\, \quar \pr_{x_2}^{\,\, 2} \big )^n 
{1\over y^{2(S+1)}} y_\nu \, . & \Ione \cr}
$$
Since
\eqn\prI{
\pr_\mu {1\over y^{2S}}I_{\mu\nu}(y) = 2(S-d+1) {1\over y^{2(S+1)}} y_\nu \, ,
}
we may therefore write for the $\ell=1$ case
\eqn\Cone{\eqalign{
C^{a,b}(s,\pr)_\mu = {}& {1\over B(a+1,b+1)} \int_0^1 \!\! \d \alpha\,
\alpha^{a}(1-\alpha)^{b}e^{\alpha s{\cdot \pr}}
\sum_{n=0} {1\over n!}
\big ( {- \alpha}(1-\alpha)\quar  s^2 \pr^{2} \big )^n \cr
&{}\times \bigg ( {1 \over (S + 1 -\half d)_n} s_\mu \cr
&\quad \ {} + {s^2 \over (S + 1 -\half d)_{n+1}}
{(a+1-\half d) \alpha - (b+1-\half d)(1-\alpha) \over 2(S-d+1)}\,\pr_\mu 
\bigg ) \, . } 
}

For $\ell=2, \ S=a+b+2$ the calculation is similar although more tedious. 
Following
the same route as led to \Irep\ we have, using \diff\ for $\ell=0,1,2$ and
for $\C_{\nu_1\nu_2}$ an arbitrary symmetric traceless tensor,
\eqn\Irept{\eqalign{\!\!\!\!\!
{1\over r_{13}^{\, \, a} \, r_{23}^{\, \, b}}&\, Z_{\nu_1}Z_{\nu_2} 
\C_{\nu_1\nu_2} \cr
= {}& {1\over S(S+1)} {1\over B(a+2,b+2)} \int_0^1 \!\! \d \alpha\,
\alpha^{a-1}(1-\alpha)^{b-1} \sum_{n=0} {1\over n!}
\big ( A\, \quar \pr_{x_2}^{\,\, 2} \big )^n {1\over y^{2S}}\C_{\nu_1\nu_2} \cr
{}\times {}&
\bigg ( {S(S-1) \over (S + 1 -\half d)_n} \,\alpha^2(1-\alpha)^2 
x_{12\nu_1}x_{12\nu_2}\cr
& {} + {2(S-1)\over (S -\half d)_n}\, \alpha(1-\alpha)
\big ( (b+1)\alpha - (a+1) (1-\alpha)\big ) 
x_{12\nu_1} y_{\nu_2} \cr
& {} + {1\over (S-1 -\half d)_n} \big ( b(b+1)\alpha^2 + a (a+1) (1-\alpha)^2
- 2 (b+1)(a+1)\alpha(1-\alpha) \big ) y_{\nu_1} y_{\nu_2} \bigg ) \, , \cr}
}
We may now write
\eqn\SSrr{\eqalign{
{S-1\over S+1}\, {1\over y^{2S}} x_{12\nu_1} x_{12\nu_1} = {}&
{1\over y^{2S}}\, x_{12\mu_1} x_{12\mu_1} I_{\mu_1\nu_1}(y) I_{\mu_2\nu_2}(y)\cr
&{} - {S-1\over S(S+1)}\, {\d \over \d \alpha} {1\over y^{2S}} 
( x_{12\nu_1} y_{\nu_2} + x_{12\nu_2} y_{\nu_1} ) \cr 
&{} - {1\over S(S+1)}\, {\d^2 \over \d \alpha^2} 
{1\over y^{2S}}y_{\nu_1} y_{\nu_2} 
- {2\over S+1}\, {r_{12}\over  y^{2(S+1)}}y_{\nu_1} y_{\nu_2} \, . \cr}
}
The resulting expression has three pieces, the first of which comes from 
the first line of \SSrr\ in \Irept\ and is readily seen to be
\eqn\Itwoa{\eqalign{
\!\!\!\!\!\!\!\! {1\over B(a+2,b+2)}& \int_0^1 \!\! \d \alpha\,
\alpha^{a+1}(1-\alpha)^{b+1} \cr
\times {}& \sum_{n=0} {1\over n!}{1\over (S + 1 -\half d)_n}
\big ( A\, \quar \pr_{x_2}^{\,\, 2} \big )^n {1\over y^{2S}}
x_{12\mu_1} x_{12\mu_1} I_{\mu_1\nu_1}(y) I_{\mu_2\nu_2}(y)\C_{\nu_1\nu_2} 
 \, . \cr}
}
After integrating by parts the remaining terms in \SSrr\ then in addition
to \Itwoa\ we have
\eqnn\JJ
$$\eqalignno{\!\!\!\!\!
{}& {S-1\over S+1} {2r_{12} \over B(a+2,b+2)} \int_0^1 \!\! \d \alpha\,
\alpha^{a+1}(1-\alpha)^{b+1} 
\big ( (a+1-\half d) \alpha - (b+1-\half d)(1-\alpha) \big ) \cr
& \qquad \qquad \qquad \qquad \quad {} \times \sum_{n=0} {1\over n!}
{1 \over (S + 1 -\half d)_{n+1}}\, \big ( A\, \quar \pr_{x_2}^{\,\, 2} \big )^n
{1\over y^{2(S+1)}} x_{12\nu_1} y_{\nu_2} \C_{\nu_1\nu_2} \cr
&{}+ {1\over S+1} {r_{12} \over B(a+2,b+2)} \int_0^1 \!\! \d \alpha\,
\alpha^{a}(1-\alpha)^{b} \cr
& \qquad \  {} \times \sum_{n=0} {1\over n!}
\bigg ( {1 \over (S -\half d)_{n+2}} J - {2\over (S + 1 -\half d)_n}
\alpha (1-\alpha) \bigg ) \big ( A\, \quar \pr_{x_2}^{\,\, 2} \big )^n
{1\over y^{2(S+1)}} y_{\nu_1} y_{\nu_2} \C_{\nu_1\nu_2} \, , \cr
&{} J = (a+1-\half d) \big ( (S+b -\half d)n + 2 (S-\half d) (b+1) \big )
\alpha^2 \cr
& \quad {}+ (b+1-\half d) \big ( (S+a -\half d)n + 2 (S-\half d) (a+1) \big )
(1-\alpha)^2\cr
& \quad {}- 2\big ( ((S-\half d) (S-\half d-1) -(a+1)(b+1) ) n\cr
& \qquad \quad {} + (S-\half d) ( (S-\half d-1)(S+1) - 2 (a+1)(b+1) )\big ) 
\alpha (1-\alpha) \, . & \JJ \cr}
$$
We may now use, similarly to \SSr,
\eqn\SSrr{\eqalign{
{S-1\over S+1} {1\over y^{2(S+1)}} x_{12\nu_1} y_{\nu_2} \C_{\nu_1\nu_2} 
= {}& {1\over y^{2(S+1)}}  x_{12\mu_1}
I_{\mu_1\nu_1}(y)y_{\nu_2} \C_{\nu_1\nu_2} \cr
&{} - {1\over S+1} {\d \over \d \alpha} {1\over y^{2(S+1)}} 
y_{\nu_1} y_{\nu_2} \C_{\nu_1\nu_2}  \, , \cr}
}
so that the first term in \JJ\ gives a contribution,
\eqn\Itwob{\eqalign{
\!\!\!\!\!\!\!\! {r_{12}\over B(a+2,b+2)}& \int_0^1 \!\! \d \alpha\,
\alpha^{a+1}(1-\alpha)^{b+1} 
\big ( (a+1-\half d) \alpha - (b+1-\half d)(1-\alpha) \big ) \cr
\times {}& \sum_{n=0} {1\over n!}{1\over (S + 1 -\half d)_{n+1}}
\big ( A\, \quar \pr_{x_2}^{\,\, 2} \big )^n {1\over y^{2(S+1)}}
x_{12\mu_1} I_{\mu_1\nu_1}(y) y_{\nu_2} \C_{\nu_1\nu_2} \, . \cr}
}
After another integration by parts the final contribution becomes
\eqnn\Itwoc
$$\eqalignno{
\!\!\!\!\!\!\!\! {r_{12}{}^{\! 2}\over B(a+2,b+2)}& \int_0^1 \!\! \d \alpha\,
\alpha^{a+1}(1-\alpha)^{b+1}\cr
&\quad {}\times \big ( (a+1-\half d) (a+2-\half d) \alpha^2 + 
(b+1-\half d)(b+2-\half d) (1-\alpha)^2 \cr
&\qquad \quad {} - 2 (a+1-\half d) (b+1-\half d) \alpha (1-\alpha) \big ) \cr
&{} \times \sum_{n=0} {1\over n!}{1\over (S + 1 -\half d)_{n+2}}
\big ( A\, \quar \pr_{x_2}^{\,\, 2} \big )^n {1\over y^{2(S+2)}}
y_{\nu_1}  y_{\nu_2} \C_{\nu_1\nu_2} \, . & \Itwoc \cr}
$$
Just as with \prI\ we may now use
\eqn\prII{\eqalign{
\pr_{\mu_2} {1\over y^{2S}}I_{\mu_1\nu_1}(y) I_{\mu_2\nu_2}(y)\C_{\nu_1\nu_2}
= {}& 
2(S-d) {1\over y^{2(S+1)}} I_{\mu_1\nu_1}(y) y_{\nu_2} \C_{\nu_1\nu_2} \, ,\cr
\pr_{\mu_1} \pr_{\mu_2} {1\over y^{2S}}I_{\mu_1\nu_1}(y) I_{\mu_2\nu_2}(y)
\C_{\nu_1\nu_2}
= {}& 2(S-d)(S-d+1){1\over y^{2(S+1)}} y_{\nu_1}  y_{\nu_2} \C_{\nu_1\nu_2}\, ,
\cr}
}
in \Itwob\ and \Itwoc\  to write the sum of \Itwoa,  \Itwob\ and \Itwoc\
in the form
\eqn\Ctwo{
C^{a,b}(x_{12},\pr_{x_2})_{\mu_1\mu_2}{1\over r_{23}^{\, S}}
I_{\mu_1\nu_1}(x_{23}) I_{\mu_2\nu_2}(x_{23})\C_{\nu_1\nu_2}\, ,
}
defining $C^{a,b}(s,\pr)$ for $\ell=2$.

Both this result and \Cone\ may be expressed in terms of the differential
operators
\eqn\Cdiff{
C^{a,b}_{\kappa}(s,\pr) = {1\over B(a,b)} \int_0^1 \!\!\! \d \alpha \, 
\alpha^{a-1}(1-\alpha)^{b-1}e^{\alpha s{\cdot \pr}}
\sum_{n=0}{1\over n!(\kappa)_n}
\big ( {- \alpha(1-\alpha)\quar s^2 \pr^2} \big )^n \, .
}
For any $\ell$, neglecting terms ${\rm O}(s^2)$, 
$C^{a,b}(s,\pr){\cdot \C}$ has the form
$C^{a+\ell,b+\ell}_{S+1-{1\over 2}d}(s,\pr)s_{\mu_1} \dots s_{\mu_\ell}
\C_{\mu_1 \dots \mu_\ell}$.

\appendix{B}{Calculation for $\ell=1$}

Here we consider \defGl\ for $\ell=1$,
\eqn\Glone{
C^{a,b}(x_{12}, \pr_{x_2})_\mu{1\over 
r_{23}^{\,\,e}\, r_{24}^{\,\,f}}Y_\mu
= {1\over r_{14}^{\,\,a}\, r_{24}^{\,\,b}}
\bigg ({r_{14}\over r_{13}} \bigg )^{\! e}
G^{(1)} (b,e,S;u,v) \, ,
}
with $C^{a,b}(x_{12}, \pr_{x_2})_\mu$ given by  \Cone, and evaluate directly
$G^{(1)} (b,e,S;u,v)$ following a 
similar route to that described for $\ell=0$ in \Pet\ and sketched in \one.
Using an integral representation similar to \Irep\ the first term \Cone\
leads to a contribution,
\eqn\Ione{\eqalign{
{1\over S}& {1\over B(a+1,b+1)B(e+1,f+1)} \int_0^1 \!\! \d \alpha\,
\alpha^{a}(1-\alpha)^{b} \int_0^1 \!\! \d \beta\, \beta^{e-1}(1-\beta)^{f-1}\cr
&{}\times \sum_{m,n=0} {1\over m!n!} {(S)_n \over (S+1-\half d)_m} 
A^m B^n \big ( \quar  z^2 \pr_z^{2} \big )^m \cr
&\ \ {}\times {1\over z^{2(S+n)}}\,  x_{12} \! \cdot \!
\Big ( (S-1) \beta(1-\beta) x_{34} + \big ( e (1-\beta ) - f\beta \big )
z \Big ) \, , \cr}
}
for
\eqn\zAB{
z= x_{24} + \alpha x_{12} - \beta x_{34} \, , \qquad
A=-\alpha(1-\alpha) r_{12} \, , \quad B=-\beta(1-\beta ) r_{34} \, .
}
Carrying out the differentiation according to \diff\ and using
\eqn\Sz{
2(S+m+n-1){1\over z^{2(S+m+n)}} \, x_{12}{\cdot z} = -{\d \over \d \alpha}
{1\over z^{2(S+m+n-1)}}\, , }
allows \Ione\ to be rewritten as
\eqnn\Ionea
$$\eqalignno{
{S-1\over S}& {x_{12}{\cdot x_{34}}
\over B(a+1,b+1)B(e+1,f+1)} \int_0^1 \!\! \d \alpha\,
\alpha^{a}(1-\alpha)^{b} \int_0^1 \!\! \d \beta\, \beta^{e}(1-\beta)^{f}\cr
&{}\times \sum_{m,n=0} {1\over m!n!} {(S)_{m+n}(S+1-\half d)_{m+n}
\over (S+1-\half d)_m (S+1-\half d)_n} \,{A^m B^n \over z^{2(S+m+n)}} \cr
{}+ {1\over 2S(S-1)}& {1\over B(a+1,b+1)B(e+1,f+1)} \int_0^1 \!\! \d \alpha\,
\alpha^{a-1}(1-\alpha)^{b-1} \int_0^1 \!\! \d \beta\, 
\beta^{e-1}(1-\beta)^{f-1}\cr
& \qquad \qquad \qquad \qquad \qquad 
{} \times \big ( a (1-\alpha) - b \alpha \big)
\big ( e (1-\beta ) - f\beta \big ) \cr
&{}\times \sum_{m,n=0} {1\over m!n!} {(S-1)_{m+n}(S-\half d)_{m+n} 
\over (S-\half d)_m (S-\half d)_n} \,{A^m B^n \over z^{2(S-1+m+n)}} \cr
{}+ {1\over 2S}& {r_{12}\over B(a+1,b+1)B(e+1,f+1)} \int_0^1 \!\! \d \alpha\,
\alpha^{a}(1-\alpha)^{b} \int_0^1 \!\! \d \beta\, 
\beta^{e-1}(1-\beta)^{f-1}\cr
& \qquad \qquad \qquad \qquad 
{} \times \big ( (a-\half d +1 )\alpha - (b - \half d +1 )(1-\alpha)\big)
\big ( e (1-\beta ) - f\beta \big ) \cr
&{}\times \sum_{m,n=0} {1\over m!n!} {(S)_{m+n}(S-\half d)_{m+n+1}    
\over (S-\half d)_{m+2} (S-\half d)_n} \,{A^m B^n \over z^{2(S+m+n)}} \, .
& \Ionea \cr}
$$
The term in \Cone\ involving $\pr_\mu$ also gives a contribution
\eqn\Itwo{\eqalign{
- {1\over 2S}& {r_{12}\over B(a+1,b+1)B(e+1,f+1)} \int_0^1 \!\! \d \alpha\,
\alpha^{a}(1-\alpha)^{b} \int_0^1 \!\! \d \beta\,
\beta^{e-1}(1-\beta)^{f-1}\cr
& \qquad \qquad \qquad \qquad
{} \times \big ( (a-\half d +1 )\alpha - (b - \half d +1 )(1-\alpha)\big)
\big ( e (1-\beta ) - f\beta \big ) \cr
&{}\times \sum_{m,n=0} {1\over m!n!} {(S)_{m+n}(S+1-\half d)_{m+n}
\over (S+1-\half d)_{m+1} (S+1-\half d)_n} \,{A^m B^n \over z^{2(S+m+n)}}\cr
{}+ {1\over 2}& {e-f\over S-d+1}
{r_{12}\,r_{34}\over B(a+1,b+1)B(e+1,f+1)} \int_0^1 \!\! \d \alpha\,
\alpha^{a}(1-\alpha)^{b} \int_0^1 \!\! \d \beta\,
\beta^{e}(1-\beta)^{f}\cr
& \qquad \qquad \qquad \qquad \qquad
{} \times \big ( (a-\half d +1 )\alpha - (b - \half d +1 )(1-\alpha)\big)\cr
&{}\times \sum_{m,n=0} {1\over m!n!} {(S+1)_{m+n}(S+2-\half d)_{m+n}
\over (S+1-\half d)_{m+1} (S+2-\half d)_n} \,{A^m B^n \over z^{2(S+1+m+n)}}
\, . }
}

To obtain the form shown in \Glone\ requires three critical steps. First,
writing $z^2 = A + B +C$ where
\eqn\defC{
C=\alpha(1-\beta)r_{14} + \beta(1-\alpha)r_{23} + \alpha\beta r_{13}
+ (1-\alpha)(1-\beta)r_{24} \, ,
}
we use
\eqn\summn{
\sum_{m,n=0} {1\over m!n!} {(\lambda)_{m+n} (\kappa)_{m+n} \over
(\kappa)_m (\kappa)_n} \, {A^m B^n \over z^{2(\lambda+m+n)}} =
{1\over C^\lambda} \sum_{m=0} {(\lambda)_{2m}  \over m! (\kappa)_m}
\, \Big ( {AB\over C^2} \Big )^{\! m} \, .
}
Secondly we require the $\alpha,\beta$ integrals to be of the form
\eqn\ist{\eqalign{
\int_0^1 \!\!\! \d \alpha \, \alpha^{a-1}(1-\alpha)^{b-1}{1\over C^{a+b}}
={}& B(a,b) \, {1\over s^a \, t^b} \, , \cr
s={}& \beta r_{13}+(1-\beta)r_{14} \, , \ t=\beta r_{23}+(1-\beta)r_{24} \, ,\cr
\int_0^1 \!\!\! \d \beta \, \beta^{e-1}(1-\beta)^{f-1}{1\over C^{e+f}}
={}& B(e,f) \, {1\over \hs^e \, \hht^f} \, , \cr
\hs={}& \alpha r_{13}+(1-\alpha)r_{23} \, , \ 
\hht =\alpha r_{14}+(1-\alpha)r_{24} \, ,\cr}
}
and then finally, for $\lambda=a+b=e+f$,
\eqn\istt{\eqalign{
\int_0^1 \!\!\! \d \beta \, \beta^{e-1}(1-\beta)^{f-1} \, {1\over s^a \, t^b}
= {}& B(e,f) \, {1\over r_{14}^{\,\,a}\, r_{24}^{\,\,b}}
\bigg ({r_{14}\over r_{13}} \bigg )^{\! e}F(b,e;\lambda;1-v) \, , \cr
\int_0^1 \!\!\! \d \alpha \, \alpha^{a-1}(1-\alpha)^{b-1} \,
{1\over \hs^e \, \hht^f} = {}& B(a,b) \, 
{1\over r_{14}^{\,\,a}\, r_{24}^{\,\,b}}
\bigg ({r_{14}\over r_{13}} \bigg )^{\! e} F(b,e;\lambda;1-v) \, . \cr}
}

To apply these results to \Ionea\ and \Itwo\ requires some manipulation. We
may immediately apply \summn\ to the first two terms in \Ionea. In the
first term we write
\eqn\abxx{
2\alpha\beta\, x_{12}{\cdot x_{34}} = - C + r_{24} +
\alpha (r_{14}-r_{24}) + \beta (r_{23} - r_{24}) \, ,
}
and the $-C$ piece may be combined with the second term in \Ionea, using
$a (1-\alpha) - b \alpha = (S-1)(1-\alpha) - b, \  e (1-\beta ) - f\beta 
= (S-1) (1-\beta ) - f$, so that either the $\alpha$ or $\beta$ integration
may be carried out using \ist. After further algebra these two terms
become
\eqnn\Ioneb
$$\eqalignno{
{1\over r_{14}^{\,\,a}\, r_{24}^{\,\,b}}&
\bigg ({r_{14}\over r_{13}} \bigg )^{\! e} \, {S(S-1)\over 2ae} \big (
G(b,e,S-\half d,S-1;u,1-v) \cr
& \qquad \qquad \qquad \qquad {} - G(b+1,e,S+1-\half d,S;u,1-v) \big ) \cr
{}- {}& \half (S-1){r_{12}\,r_{34}\over B(a+1,b+1)B(e+1,f+1)} 
\int_0^1 \!\! \d \alpha\,
\alpha^{a}(1-\alpha)^{b+1} \int_0^1 \!\! \d \beta\,
\beta^{e}(1-\beta)^{f+1}\cr 
&\qquad \qquad {}\times (S-d+1) {1\over C^{S+1}} \sum_{m=0} {1\over m!}
{(S+1)_{2m}\over (S-\half d)_{m+2}} \, \Big ( {AB\over C^2} \Big )^{\! m} \cr
{} + {}& r_{12}\, r_{34}\, {S-1\over 2a} \, {a-\half d +1 \over
B(e+1,f+1)} \int_0^1 \!\!\! \d \beta \, \beta^{e}(1-\beta)^{f+1} \, 
{1\over s^{a+1} \, t^{b+1}} \cr
& \qquad \qquad \qquad {}\times \sum_{m=0} {1\over m!} 
{(a)_{m+1}(b+1)_m\over (S-\half d)_{m+2}} \Big ( {r_{12}r_{34} \beta
(1-\beta) \over s\, t} \Big )^{\! m} \cr
{} + {}& r_{12}\, r_{34}\, {S-1\over 2e} \, {e-\half d +1 \over
B(a+1,b+1)} \int_0^1 \!\!\! \d \alpha \, \alpha^{a}(1-\alpha)^{b+1} \, 
{1\over \hs^{a+1} \, \hht^{b+1}} \cr
& \qquad \qquad \qquad {}\times \sum_{m=0} {1\over m!} 
{(e)_{m+1}(f+1)_m\over (S-\half d)_{m+2}} \Big ( {r_{12}r_{34} \alpha
(1-\alpha) \over \hs\, \hht} \Big )^{\! m} \, , & \Ioneb \cr}
$$
with $G$ defined by the series in \defG. Combining the last term in
\Ionea\ with \Itwo\ leads to
\eqnn\Itwoa
$$\eqalignno{
{1\over 2}& {S-1\over S-d+1}
{r_{12}\,r_{34}\over B(a+1,b+1)B(e+1,f+1)} \int_0^1 \!\! \d \alpha\,
\alpha^{a}(1-\alpha)^{b} \int_0^1 \!\! \d \beta\,
\beta^{e}(1-\beta)^{f}\cr
& {} \times \big ( (a-\half d +1 )\alpha - (b - \half d +1 )(1-\alpha)\big)
\big ( (e-\half d +1 )\beta - (f - \half d +1 )(1-\beta)\big) \cr
&{}\times \sum_{m,n=0} {1\over m!n!} {(S+1)_{m+n}(S-\half d)_{m+n+2}
\over (S-\half d)_{m+2} (S-\half d)_{n+2}} \,{A^m B^n \over z^{2(S+1+m+n)}}
\, . & \Itwoa \cr}
$$
This is symmetric and the summation formula
\summn\ may now be applied and the result combined
with the corresponding term in \Ioneb\ using
\eqn\abS{\eqalign{
& \big ( (a-\half d +1 )\alpha - (b - \half d +1 )(1-\alpha)\big)
\big ( (e-\half d +1 )\beta - (f - \half d +1 )(1-\beta)\big)\cr 
&\quad {} - (S-d+1)^2 (1-\alpha)(1-\beta) \cr
&{}= (a-\half d +1 )(e - \half d +1 ) \cr
&\quad {} - (S-d+1) 
\big ( (a-\half d +1 )(1-\beta) + (e - \half d +1 )(1-\alpha)\big) \, . \cr}
}
The integrals arising from the terms in the last line
involving $1-\alpha, \, 1-\beta$ then cancel the remaining 
$\alpha, \beta$ integrals in \Ioneb\ leaving just the first term
in \abS\ for which the associated integral may be evaluated giving
\eqn\Itwob{\eqalign{
{1\over r_{14}^{\,\,a}\, r_{24}^{\,\,b}}
\bigg ({r_{14}\over r_{13}} \bigg )^{\! e}& \, 
\half (S-1) {(a-\half d +1 ) (e - \half d +1 )\over (S-d+1)
(S-\half d)(S-\half d +1)}\cr
&{}\times  u G(b+1,e+1,S-\half d+2,S+1;u,1-v) \, .\cr}
}
In consequence from \Ioneb\ and \Itwob\ we have finally altogether
\eqn\Gone{\eqalign{
\!\!\!\!\!\!\! & G^{(1)} (b,e,S;u,v) \cr
\!\!\!\!\!\!\! &{} = {S(S-1)\over 2ae} \big ( G(b,e,S-\half d,S-1;u,1-v) 
- G(b+1,e,S+1-\half d,S;u,1-v) \big ) \cr
\!\!\!\!\!\!\! &\ {}+ {(S-1)(a-\half d +1 ) (e - \half d +1 )\over 2(S-d+1)
(S-\half d)(S-\half d +1)}\,
u G(b+1,e+1,S-\half d+2,S+1;u,1-v) \, .\cr}
}

The result \Gone\ is different from that which is given by \recurG\ for
$\ell =1$ together with \Gzero. To show the equivalence of the two
expressions it is sufficient to
use the following relations for the function $G$ defined by \defG,
\eqn\relG{\eqalign{
\alpha\, & vG(\alpha+1,\beta+1,\gamma,\de+1;u,1-v)
+(\de-\alpha-\beta)G(\alpha,\beta+1,\gamma,\de+1;u,1-v) \cr
&\quad {} -(\de-\beta)G(\alpha,\beta,\gamma,\de+1;u,1-v) \cr
&{}= {\alpha(\de-\beta)(\de-\alpha-\beta)\over \gamma(\de+1)}\,
u G(\alpha+1,\beta+1,\gamma+1,\de+2;u,1-v) \, , \cr
\beta & G(\alpha,\beta+1,\gamma,\de+1;u,1-v) +(\de-\beta)
G(\alpha,\beta,\gamma,\de+1;u,1-v) \cr
&\quad {} -\de G(\alpha,\beta,\gamma-1,\de;u,1-v) \cr
&{}= -(\de-\alpha-\gamma+1)
{\alpha\beta(\de-\beta)\over (\gamma-1) \gamma(\de+1)}\,
u G(\alpha+1,\beta+1,\gamma+1,\de+2;u,1-v) \, , \cr
\alpha & G(\alpha+1,\beta,\gamma,\de+1;u,1-v) - \beta
G(\alpha,\beta+1,\gamma,\de+1;u,1-v) \cr
&\quad {} -(\alpha-\beta) G(\alpha,\beta,\gamma,\de+1;u,1-v) \cr
&{}= -(\alpha-\beta)
{\alpha\beta\over \gamma(\de+1)}\,
u G(\alpha+1,\beta+1,\gamma+1,\de+2;u,1-v) \, . \cr}
}
These may be obtained from relations given in \one\ which
express $G(\alpha,\beta,\gamma,\de;u,1-v)$ in terms
of $G(\alpha',\beta',\gamma',\de';u,1-v)$ with $\de'-\gamma'
= \de -\gamma+1$.

\appendix{C}{Identities for $H$}

{}From the definition \defH\ and properties of the function $G$ the following
relations were obtained in \one,
\eqna\Hrel
$$\eqalignno{
H(\alpha,\beta ,\gamma,\delta;u,v) = {}& v^{-\alpha}
H(\alpha,\de-\beta,\gamma,\delta;u/v,1/v) & \Hrel a\cr
= {}& v^{\de-\alpha-\beta}
H(\de-\alpha,\de-\beta,\gamma,\delta;u,v) & \Hrel b\cr
={}& H(\alpha,\beta,\alpha+\beta-\de+1,\alpha+\beta-\gamma+1;v,u) & \Hrel c\cr
={}& u^{1-\gamma}
H(\alpha-\gamma+1,\beta-\gamma+1,2-\gamma,\delta-2\gamma+2;u,v)\, .& \Hrel d\cr}
$$
In terms of the result \DfourH\ for  the four point function these give
\eqn\Dsym{\eqalign{
\oD_{\Delta_1\, \Delta_2\, \Delta_3\, \Delta_4}(u,v) = {}&
\oD_{\Sigma{-\Delta_3}\,\Sigma{-\Delta_4}\,\Sigma{-\Delta_1}\,
\Sigma{-\Delta_2}}(u,v) \cr
={}& v^{-\Delta_2}\oD_{\Delta_1\, \Delta_2\, \Delta_4\, \Delta_3}(u/v,1/v) \cr
={}& v^{\Delta_4-\Sigma} \,
\oD_{\Delta_2\, \Delta_1\, \Delta_3\, \Delta_4}(u/v,1/v) \cr
={}& v^{\Delta_1+\Delta_4-\Sigma} \,
\oD_{\Delta_2\, \Delta_1\, \Delta_4\, \Delta_3}(u,v) \cr
={}& \oD_{\Delta_3\, \Delta_2\, \Delta_1\, \Delta_4}(v,u) \cr
={}& u^{\Delta_3+\Delta_4-\Sigma} \,
\oD_{\Delta_4\, \Delta_3\, \Delta_2\, \Delta_1}(u,v) \, , \cr} 
}
which reflect the symmetry under the interchanges $x_i,\Delta_i \leftrightarrow
x_j,\Delta_j$  for various $i,j$.

Results obtained in \one\ for $G$ also imply
\eqnn\relH
$$\eqalignno{
&(\alpha-\beta)H(\alpha,\beta,\gamma,\delta;u,v) =
H(\alpha+1,\beta,\gamma,\delta+1;u,v) - 
H(\alpha,\beta+1,\gamma,\delta+1;u,v) \, , \cr
& (\de-\alpha-\beta)H(\alpha,\beta,\gamma,\delta;u,v) =
H(\alpha,\beta,\gamma,\delta+1;u,v) -
v H(\alpha+1,\beta+1,\gamma,\delta+1;u,v) \, ,\cr
& (1-\gamma)H(\alpha,\beta,\gamma,\delta;u,v) = 
H(\alpha,\beta,\gamma-1,\delta;u,v) 
- u H(\alpha+1,\beta+1,\gamma+1,\delta+2;u,v) \, , \cr 
&(\de - \gamma - \beta + 1) H(\alpha,\beta,\gamma,\delta;u,v) =
H(\alpha,\beta,\gamma-1,\delta;u,v) + H(\alpha+1,\beta,\gamma,\delta+1;u,v) \cr
& \qquad \qquad \qquad \qquad \qquad \qquad \qquad \qquad
{}+ H(\alpha,\beta,\gamma,\delta+1;u,v) \, . & \relH \cr}
$$
The first three relations imply
\eqn\relD{\eqalign{
(\Delta_2 + \Delta_4 - \Sigma)
\oD_{\Delta_1\, \Delta_2\, \Delta_3\, \Delta_4}(u,v) ={}&
\oD_{\Delta_1\, \Delta_2{+1}\, \Delta_3\, \Delta_4{+1}}(u,v) -
\oD_{\Delta_1{+1}\, \Delta_2\, \Delta_3{+1}\, \Delta_4}(u,v) \, , \cr
(\Delta_1 + \Delta_4 - \Sigma)
\oD_{\Delta_1\, \Delta_2\, \Delta_3\, \Delta_4}(u,v) ={}&
\oD_{\Delta_1{+1}\, \Delta_2\, \Delta_3\, \Delta_4{+1}}(u,v) -
v \oD_{\Delta_1\, \Delta_2{+1}\, \Delta_3{+1}\, \Delta_4}(u,v) \, , \cr
(\Delta_3 + \Delta_4 - \Sigma)
\oD_{\Delta_1\, \Delta_2\, \Delta_3\, \Delta_4}(u,v) ={}&
\oD_{\Delta_1\, \Delta_2\, \Delta_3{+1}\, \Delta_4{+1}}(u,v) - u
\oD_{\Delta_1{+1}\, \Delta_2{+1}\, \Delta_3\, \Delta_4}(u,v) \, , \cr}
}
which are equivalent to results obtained in \FreeD. The last relation in
\relH\ also gives
\eqn\relDa{\eqalign{
\Delta_4 \oD_{\Delta_1\, \Delta_2\, \Delta_3\, \Delta_4}(u,v) ={}&
\oD_{\Delta_1\, \Delta_2\, \Delta_3{+1}\, \Delta_4{+1}}(u,v)
+ \oD_{\Delta_1\, \Delta_2{+1}\, \Delta_3\, \Delta_4{+1}}(u,v) \cr
&{}+ \oD_{\Delta_1{+1}\, \Delta_2\, \Delta_3\, \Delta_4{+1}}(u,v) \, .\cr }
}
We also have
\eqn\diffH{\eqalign{
\pr_v H(\alpha,\beta,\gamma,\delta;u,v) = {}& 
- H(\alpha+1,\beta+1,\gamma,\delta+1;u,v) \, , \cr
\pr_u H(\alpha,\beta,\gamma,\delta;u,v) = {}& 
- H(\alpha+1,\beta+1,\gamma+1,\delta+2;u,v) \, , \cr}
}
or
\eqn\diffD{\eqalign{
\pr_v \oD_{\Delta_1\, \Delta_2\, \Delta_3\, \Delta_4}(u,v) = {}&
- \oD_{\Delta_1\, \Delta_2{+1}\, \Delta_3{+1}\, \Delta_4}(u,v) \, \cr
\pr_u \oD_{\Delta_1\, \Delta_2\, \Delta_3\, \Delta_4}(u,v) = {}&
- \oD_{\Delta_1{+1}\, \Delta_2{+1}\, \Delta_3\, \Delta_4}(u,v) \, . \cr}
}

When $\gamma$ is an integer the definition $\defH$ gives $\ln u$ terms as
well as an expansion in powers in $u,\, 1-v$. For $\gamma=n=1,2,\dots$
we easily find the terms involving $\ln u$ as
\eqn\lnH{
H(\alpha,\beta,n,\delta;u,v)_{\rm log. terms}
= \ln u \, {(-1)^n \over (n-1)!}\,
{\Gamma(\alpha) \Gamma(\beta) \Gamma(\delta-\alpha) \Gamma(\delta-\beta) 
\over \Gamma(\delta)}  G(\alpha,\beta,n,\delta;u,1-v) \, .
}
If $\gamma= n= 0, -1,-2,\dots$ then the leading log term is $u^{1-n}\ln u$ and
the corresponding formula may be obtained from \lnH\
using \Hrel{d}. By virtue of \Hrel{c}\ there are
similarly terms involving $\ln v$ in an expansion for $v\sim 0$ when 
$\alpha+\beta-\de$ is an integer. For $\gamma=1$, which is of relevance
in \Dfour\ when $\Delta_1+\Delta_2=\Delta_3+\Delta_4$, a complete
formula for the additional power terms as well as the log. terms
displayed in \lnH\ is given in \one. For $\gamma=0$ the corresponding
formula is
\eqnn\defHH
$$\eqalignno{
\!\!\!\!\!\!\!\!\!\! H(\alpha,\beta,0,\delta;u,v) ={}&  {1\over \Gamma(\delta)}
\Gamma(\alpha) \Gamma(\beta) \Gamma(\delta-\alpha) \Gamma(\delta-\beta) \,
\bigg \{ F(\alpha,\beta;\delta;1-v) \cr
{}& - \!\!\!\! \sum_{m=1,n=0} {(\delta-\alpha)_m
(\delta-\beta)_m \over (m-1)! \, m!} \,
{(\alpha)_{m+n} (\beta)_{m+n} \over n! \, (\delta)_{2m+n}} 
\big ( {-\ln u} + f_{mn} \big )
u^m (1-v)^n \bigg \} \, , \cr
f_{mn} = {}& \psi(1+m) +\psi (m) + 2\psi(\de+2m+n)
- \psi(\de-\alpha+m) -\psi(\de-\beta+m) \cr
{}& - \psi(\alpha+m+n) - \psi(\beta+m+n) \, . & \defHH \cr}
$$

With the aid of the above relations we may determine $\oD_{n_1 n_2 n_3 n_4}
(u,v)$ for $n_i=1,2,\dots $, $\sum_i n_i$
even, in terms of $\oD_{1111}(u,v)$. For example if
$n_i =n=1,2,\dots$ then, from \DfourH, $\oD_{nnnn}(u,v)=H(n,n,1,2n;u,v)$,
where
\eqn\HDn{
H(n,n,1,2n;u,v) = v^{-n}H(n,n,1,2n;u/v,1/v) = H(n,n,1,2n;v,u) \, ,
}
and using \diffH\ and \Hrel{b,d}\ we may obtain the recurrence relation
\eqn\Hup{
H(n+1,n+1,1,2n+2;u,v) = \pr_u u \pr_u H(n,n,1,2n;u,v)
= \pr_v v \pr_v H(n,n,1,2n;u,v) \, .
}
For the starting point $\oD_{1111}$, with the definitions in \uvxz\ for
$z,x$ (note that $(z-x)^2 = 1+u^2+v^2-2u-2v-2uv$), we have
\eqn\Hone{
\oD_{1111}(u,v)= H(1,1,1,2;u,v) = {1\over z-x} \Phi(z,x) \, ,
}
with
\eqn\defPhi{
\Phi(z,x) = \ln zx \, \ln {1-z \over 1-x} - 
2{\rm Li}_2 (x) + 2{\rm Li}_2(z) \, ,
}
where ${\rm Li}_2$ is the dilogarithm function. The result \Hone\ with
\defPhi\ was derived in \one\ and is equivalent to results \Uss\ known for some
time for the integral in \fourp\ with $\alpha_i=1, \ d=4$. The function $\Phi$
in \defPhi\ satisfies the following critical identities
\eqn\symP{
\Phi(z,x) = - \Phi(x,z) = - \Phi(1-z,1-x) = - \Phi(z',x') \, , \qquad
x' = {x\over x-1} \, , \quad z' = {z\over z-1} \, ,
}
which depend on standard results for the dilogarithm function. The
results \symP\ are required in order to satisfy \HDn\ for $n=1$.

For $n=2$ we may use
\eqn\diffP{
\pr_u \Phi(z,x) = {1\over z-x} \Big ( {1-u-v \over u} \ln v + 2 \ln u \Big ) 
\, , \quad \pr_v \Phi(z,x) = {1\over z-x} 
\Big ( {1-u-v \over v} \ln u + 2 \ln v \Big ) \, ,
}
to obtain
\eqn\HDDs{\eqalign{
H(2,2,1,4;u,v) = {}& \Big ( {12uv \over (z-x)^5} + {1+u+v \over (z-x)^3}\Big )
\Phi(z,x) \cr
&{}+ {6\over (z-x)^4} \big ( (1+u-v)v\ln v + (1-u+v) u \ln u \big ) \cr
&{} + {2\over (z-x)^2} \big ( \ln uv + 1 \big ) \, . \cr}
}
Also for $n=3$ we have
\eqnn\HDDDs
$$\eqalignno{
H(3,3,1,6;u,v) = {}& \bigg ( {1680u^2v^2 \over (z-x)^9} 
+ \Big ( {240uv \over (z-x)^7}+ {24 \over (z-x)^5}\Big )(1+u+v) 
+  {4 \over (z-x)^3}\bigg ) \Phi(z,x) \cr
&{}+ \bigg ( \Big ( {840u\over (z-x)^8} + {100 \over (z-x)^6} \Big )
v^2 (1+u-v)  + {480uv\over (z-x)^6} \cr
&\qquad {}+ {1\over (z-x)^4} \big ( 12(1+u)
+ 76 v \big ) \bigg ) \ln v  + u \leftrightarrow v \cr
&{} + {260uv\over (z-x)^6} + {26\over (z-x)^4} ( 1+u+v) \, . & \HDDDs \cr}
$$

For recent applications \Frol\ it is necessary to know 
$D_{\Delta_1\Delta_2\Delta_3\Delta_4}$ for other small integer values
of $\Delta_i$. To this end we first determine
\eqn\HD{\eqalign{
H(1,1,1,3;u,v) ={}&  v^{-1}H(1,2,1,3;u/v,1/v) = v H(2,2,1,3;u,v) \cr
= {}& v H(2,2,2,4;v,u) = H(1,1,0,2;v,u) \, , \cr}
}
which follow from \Hrel, and correspond to $\oD_{2112}(u,v)=v^{-1}
\oD_{2121}(u/v,1/v) =v\oD_{1221}(u,v)=v\oD_{2211}(v,u)=\oD_{1122}(v,u)$. 
{}From \diffH\ we may find
\eqn\HDs{
H(1,1,1,3;u,v) = - v {1+u-v \over (z-x)^3} \Phi(z,x) -
{1\over (z-x)^2}\big ( (1-u-v) \ln u + 2 v \ln v \big ) \, .
}
Similarly for $\oD_{2233}(u,v)$ it is necessary to determine
\eqn\HDDD{\eqalign{
H(2,2,1,5;u,v) ={}&  v^{-2}H(2,3,1,5;u/v,1/v) = v H(3,3,1,5;u,v) \cr
= {}& v H(3,3,2,6;v,u) = H(2,2,0,4;v,u) \, , \cr}
}
and explicitly we have
\eqn\HDDDs{\eqalign{
H(& 2,2,1,5;u,v)\cr
= & - \bigg ( {60uv^2 \over (z-x)^7}(1+u-v) + {6v\over (z-x)^5}
\big ( v(1+u-v) + 4u \big ) + {4v\over (z-x)^3} \bigg ) \Phi (z,x)\cr
&- \bigg ( {120 uv^2 \over (z-x)^6} +{2v \over (z-x)^4}
\big ( 6(1+u) + 5v \big ) \bigg ) \ln v \cr
&- \bigg ( {60 uv \over (z-x)^6}(1-u-v) + {2 \over (z-x)^4}
\big (1-u+v - 9uv -2v^2 \big) - {1\over (z-x)^2} \bigg ) \ln u \cr
& - {10v\over (z-x)^4}(1+u-v) - {2\over (z-x)^2} \, . \cr}
}

\appendix{D}{Vector Four Point Function}

We describe here a few details concerning the derivation of the result
\fourF\ where maintaining manifest conformal invariance simplifies the
calculation (a related calculation is described by Herzog \Herz). 
The disconnected graphs contributing to 
$\l {1\over 4}F^2(x_1) {1\over 4}F^2(x_2) {1\over 4}F^2(x_3) {1\over 4}F^2(x_4)\r$
are of course straightforward while the connected graphs, using \twoF, give
\eqn\III{\eqalign{
{1\over \pi^8}\,{1\over (r_{12}r_{24}r_{34}r_{13})^2}\, {1\over 2} \Big (&
\big ( {\rm tr}(I(x_{12})I(x_{24})I(x_{43})I(x_{31})) \big )^2 \cr
&{} - {\rm tr}\big (I(x_{12})I(x_{24})I(x_{43})I(x_{31}) 
I(x_{12})I(x_{24})I(x_{43})I(x_{31})\big )\Big ) \,, \cr}
}
together with two other permutations. To evaluate the traces of the
inversion tensors in \III\ we use \hughone
\eqn\IIIr{
I(x_{1i}) I(x_{ij}) I(x_{j1}) = I(X_{1(ij)}) \, , \qquad
X_{1(ij)} = {x_{i1}\over r_{1i}} - {x_{j1}\over r_{1j}} \, , \quad
X_{1(ij)}^{\,\,\, 2} = {r_{ij}\over r_{1i}\, r_{2j}} \, ,
}
where $X_{1(ij)}$ transforms as a conformal vector at $x_1$. With \IIIr
\eqn\IIII{
{\rm tr}\big (I(x_{12})I(x_{24})I(x_{43})I(x_{31})\big ) =
{\rm tr}\big (I(X_{1(24)})I(X_{1(43)})\big ) = 4\, {\big (
X_{1(24)}{\cdot X_{1(43)}}\big )^2 \over X_{1(24)}^{\,\,\, 2} \,
X_{1(43)}^{\,\,\, 2}} \, .
}
Similarly
\eqnn\IIIII
$$\eqalignno{
{} & {\rm tr}\big (I(x_{12})I(x_{24})I(x_{43})I(x_{31}) 
I(x_{12})I(x_{24})I(x_{43})I(x_{31})\big ) \cr
&{} = {\rm tr}\big (I(X_{1(24)})I(X_{1(43)})I(X_{1(24)})I(X_{1(43)})\big )\cr
&{} = {\rm tr}\big (I(X_{1(43)})I(X_{1(24)})I(X_{1(43)})\big ) - {2\over
X_{1(24)}^{\,\,\, 2}} \, X_{1(24)}{\cdot I}(X_{1(43)})I(X_{1(24)})I(X_{1(43)})
\, {\cdot X_{1(24)}} \cr
&{} = 4 \bigg ( 1 - 2\, { (X_{1(24)}{\cdot X_{1(43)}} )^2 \over 
X_{1(24)}^{\,\,\, 2} \, X_{1(43)}^{\,\,\, 2}} \bigg )^{\! 2} \, . & \IIIII \cr}
$$
Since
\eqn\XX{
2X_{1(24)}{\cdot X_{1(43)}} = {r_{23}\over r_{12}\, r_{14}} \Big (
1 - {1\over v} - {u\over v} \Big ) \,
}
we have
\eqnn\FII
$$\eqalignno{
&\big ( {\rm tr}(I(x_{12})I(x_{24})I(x_{43})I(x_{31})) \big )^2 
 - {\rm tr}\big (I(x_{12})I(x_{24})I(x_{43})I(x_{31}) 
I(x_{12})I(x_{24})I(x_{43})I(x_{31})\big ) \cr
&{} = 16 \, {\big (X_{1(24)}{\cdot X_{1(43)}}\big )^2 \over 
X_{1(24)}^{\,\,\, 2} \, X_{1(43)}^{\,\,\, 2}}  - 4 
= 4 \, {1\over u} (v-1-u)^2 - 4 \, , & \FII \cr}
$$
and \III\ becomes
\eqn\IIIf{
{1\over \pi^8}\,{2\over (r_{12}\, r_{34})^4}\, \big ( u(v-1-u)^2 - u^2\big ) \, .
} 
The other two contributions may be obtained similarly or by applying
permutations to the results \IIIf.
\listrefs
\bye